\documentclass[hyper,11pt,letterpaper]{JHEP3}
\usepackage[dvips]{epsfig}
\usepackage{amsmath,amssymb,epsf,amsfonts}
\usepackage{cite}
\usepackage{graphicx}
\usepackage{dcolumn}
\usepackage{bm}

\def\N{{\mathcal{N}}}

\def\<{\langle}
\def\th{\theta}
\def\>{\rangle}
\def\({\left (}
\def\){\right )}
\def\[{\left[}
\def\]{\right]}

\def\beq{\begin{equation}}
\def\eeq{\end{equation}}
\def\gx{g_{xx}}
\def\gt{|g_{tt}|}

\def\E{\tilde{E}}
\def\B{\tilde{B}}

\def\Ui{U^t_{~i}}
\def\Ti{\langle T^t_{~i} \rangle}
\def\At{\tilde{A}'_t}
\def\Ax{\tilde{A}'_x}
\def\Ay{\tilde{A}'_y}
\def\mi{(m^{-1})}
\def\jt{\langle J^t \rangle}
\def\jx{\langle J^x \rangle}
\def\jy{\langle J^y \rangle}
\def\ve{\varepsilon}
\def\p{\pi}                
\def\a{\alpha}
\def\s{\sigma}
\def\lam{\lambda}
\newcommand{\noi}{\noindent}
\newcommand{\bea}{\begin{eqnarray}}
\newcommand{\eea}{\end{eqnarray}}

\def\Om{{\cal{O}}_m}
\def\Omv{\langle {\cal{O}}_m \rangle}
\def\Nf{N_f}
\def\Nc{N_c}
\def\rarrow{\rightarrow}
\def\ra{\rightarrow}

\def\a{\alpha}

\def\d{\delta}

\def\lam{\lambda}
\def\m{\mu}
\def\n{\nu}
  
\def\p{\pi}                
\def\th{\theta}                   
\def\s{\sigma}                                   

\def\O{\Omega}

\title{\LARGE The Stress-Energy Tensor of Flavor Fields from AdS/CFT}

\author{Andreas Karch,\!$^1$\footnotemark[1]\,
Andy O'Bannon,\!$^{1,2}$\footnotemark[2]\,
and Ethan Thompson\!$^1$\footnotemark[3]
\\
$^1$Department of Physics, University of Washington, Seattle, WA 98195-1560
\\
$^2$Max Planck Institut f\"{u}r Physik (Werner Heisenberg Institut) \\ F\"{o}hringer Ring 6, 80805 M\"{u}nchen, Germany}

\footnotetext[1]{E-mail: \email{karch@phys.washington.edu}}
\footnotetext[2]{E-mail: \email{ahob@mppmu.mpg.de}}
\footnotetext[3]{E-mail: \email{egthomps@u.washington.edu}}

\abstract
{
We use the AdS/CFT correspondence to study the transport properties of massive $\N=2$ hypermultiplet fields in an $\N=4$ $SU(N_c)$ super-Yang-Mills theory plasma in the large $N_c$, large 't Hooft coupling limit, and in the presence of a baryon number chemical potential and external electric and magnetic fields. In particular, we compute the flavor fields' contribution to the stress-energy tensor. We find infrared divergences in the stress-energy tensor, arising from the flavor fields' constant rate of energy and momentum loss. We regulate these divergences and extract the energy and momentum loss rates from the divergent terms. We also check our result in various limits in which the divergences are absent. The supergravity dual is a system of D7-branes, with a particular configuration of worldvolume fields, probing an AdS-Schwarzschild background. The supergravity calculation amounts to computing the stress-energy tensor of the D7-branes.
}
\preprint{MPP-2008-166 \\ INT PUB 08-49}

\begin{document}

\section{Introduction}

The conductivity tensor $\s_{ij}$ measures the electrical response of a conducting medium to externally applied fields. It is defined by
\beq
\< J_i \> = \s_{ij} \, E_j \nonumber
\eeq
where $E$ are externally applied electric fields and $\< J_i \>$ are the electrical currents induced in the medium. Similarly, the thermoelectric conductivity tensor $\alpha_{ij}$ measures the thermal response. It is defined as
\beq
\< Q_i \> = \alpha_{ij} \, E_j \nonumber
\eeq
where $\< Q_i \>$ are heat currents induced in the medium,
\beq
\< Q_i \> = \< T^{t}_{~i} \> - \mu \< J_i \>, \nonumber
\eeq
where $\< T^i_{~j} \>$ are the components of the stress-energy tensor, so that $\< T^t_{~i} \>$ are momentum densities, and $\mu$ is the chemical potential. $\alpha_{ij}$ governs the Peltier, Seebeck and Nernst effects.

In this paper we will use the the anti-de Sitter / Conformal Field Theory correspondence (AdS/CFT) \cite{Maldacena:1997re,Gubser:1998bc,Witten:1998qj} to study the heat currents that arise in response to external fields in a strongly-coupled system. As with other calculations of transport phenomena via AdS/CFT (for example the calculation of viscosity in ref. \cite{Policastro:2001yc}) the over-arching objective is to develop solvable toy models that may provide insight into the real-time behavior of strongly-correlated systems.

High-$T_c$ superconductors tend to exhibit an unusually large Nernst effect even outside the superconducting phase, which may give some hint as to the nature of the still-mysterious pairing mechanism at work in these materials. While an AdS/CFT description of high-T$_c$ superconductivity has not yet been found, AdS/CFT techniques can clarify what sort of theories could account for a large Nernst signal. For example, one potential explanation was suggested in ref. \cite{Hartnoll:2007ih}: a quantum critical point. The physics of systems near their quantum critical points is described by a strongly-coupled CFT--- a class of theories for which AdS/CFT provides many examples. Based on this logic, an AdS/CFT calculation of thermo-electric response in a strongly-coupled (2+1)-dimensional CFT was performed in ref. \cite{Hartnoll:2007ih}.

We will focus on a (3+1)-dimensional ${\cal N }=4$ supersymmetric $SU(N_c)$ Yang-Mills (SYM) theory plasma at temperature $T$. We work in the limits $\Nc \rarrow \infty$ with the 't Hooft coupling $\lambda \equiv g_{YM}^2 \Nc$ fixed, and with $\lambda \gg 1$. We will introduce a number $\Nf$ of massive $\N = 2$ supersymmetric hypermultiplets transforming in the fundamental representation of the gauge group (\textit{i.e.} flavor fields). We fix $\Nf$ such that $\Nf \ll \Nc$, and work to leading order in $\Nf/\Nc$. With mass-degenerate flavor fields, the theory has a global $U(N_f)$ symmetry, whose $U(1)_B$ subgroup we identify as baryon number (hence the subscript). We will study the theory with a finite $U(1)_B$ chemical potential and with external electric and magnetic fields that couple to anything carrying $U(1)_B$ charge.

AdS/CFT equates the SYM theory in the limits above with supergravity on the ten-dimensional spacetime $AdS_5 \times S^5$, where $AdS_5$ is (4+1)-dimensional anti-de Sitter space and $S^5$ is a five-sphere \cite{Maldacena:1997re}. The SYM theory in thermal equilibrium is dual to supergravity on an AdS-Schwarzschild spacetime, where the SYM theory temperature $T$ is identified with the Hawking temperature of the AdS-Schwarzschild black hole \cite{Witten:1998zw,Gubser:1996de}. The $\Nf$ hypermultiplets appear in the supergravity description as a number $\Nf$ of D7-branes embedded in the AdS-Schwarzschild background \cite{Karch:2002sh}. With $\Nf \ll \Nc$ D7-branes, we may treat the branes as probes and neglect their back-reaction on the geometry.  We will explain the supergravity description of the hypermultiplet mass, the $U(1)_B$ chemical potential, and the electric and magnetic fields in the sequel. Though we focus on this system, our analysis easily extends to other probe D-brane systems.

One drawback of the system studied in ref. \cite{Hartnoll:2007ih} is that it has translation invariance, which implies momentum conservation. The system thus has no way to dissipate momentum, and so the DC transport behavior was singular. For example, the DC conductivity at finite density is infinite because the charge carriers, in the presence of an external electric field but without frictional forces, accelerate forever. In ref. \cite{Hartnoll:2007ih}, this problem was addressed by introducing (by hand) a relaxation time $\tau$, representing the presence of defects. A calculation of $\tau$ in an AdS/CFT toy model (a not-quite-consistent truncation of M-theory) was later performed in ref. \cite{Hartnoll:2008hs}. A calculation of the Nernst effect entirely within the framework of a ``complete'' AdS/CFT system remains to be done.

The D7-brane probe of ref. \cite{Karch:2002sh} provides an elegant solution to the problem of singular DC transport in a translationally-invariant finite density system \cite{Karch:2007pd,O'Bannon:2007in}. The key fact in the SYM theory is that the free energy of the flavor fields and the adjoint fields have different scalings with powers of $\Nc$. The flavor fields have order $N_f N_c$ degrees of freedom, whereas the adjoint fields have order $N_c^2$. As an external electric field accelerates the flavor charge carriers, their momenta increases at a rate of order $N_c$. In a stationary state (as constructed in refs. \cite{Karch:2007pd,O'Bannon:2007in}), this external force is balanced by a dissipative force which transfers momentum to the adjoint degrees of freedom of the SYM plasma at a rate of order $N_c$. As the energy and momentum densities of the adjoint degrees of freedom are of order $N_c^2$, they can absorb the order $N_c$ contribution without experiencing any significant back-reaction up to times of order $N_c$. That is, the adjoint degrees of freedom effectively act as a heat sink into which the flavor fields can dissipate energy and momentum at a constant rate. Only at very late times (parametrically large in $\Nc$), will the adjoint degrees of freedom acquire an order-one velocity. Before that time, their motion is negligible. As a result, the DC conductivity of the flavor fields is finite
even at finite density \cite{Karch:2007pd,O'Bannon:2007in}.

As for the thermo-electric response, we will see below that the issue of dissipation becomes more subtle than it was for the conductivity. In particular, no meaningful separation exists between the flavor fields' and the adjoint fields' contributions to the heat currents.  We must face the fact that if external work is done on the system then the heat current is not stationary. Nevertheless, we are able to extract many interesting pieces of information from the momentum and heat currents. For generic background fields, we calculate the rate of change of the heat and momentum currents (which we, irrespective of the sign, will refer to as ``loss rates'') and find they agree with field theory expectations. We also study special limits, or carefully crafted observables, that are insensitive to the loss rates.

The general strategy is to match conserved charges in the bulk of AdS with those in the SYM theory. Energy and momentum densities in the bulk map one-to-one to the corresponding densities in the SYM theory. This strategy has been implicit in much recent work involving fundamental strings in the bulk which are thought to describe states with large energy or angular momentum, starting with ref. \cite{Gubser:2002tv}. Here we generalize this idea to probe D-branes. We find that in general this strategy sometimes misses important contributions to the energy of the state which come from the back-reaction of the probe on the geometry, and in particular the change in the horizon of the AdS-Schwarzschild black hole due to the probe. Based on these observations, we can clarify a few important points in the literature. Momentum currents, on the other hand, can reliably be calculated in the probe approximation.

This paper is organized as follows. In section \ref{chargessection} we review the basic strategy and investigate the issue of back-reaction with a few illustrative examples. In the remainder of the paper we apply the strategy to the probe D7-brane. In section \ref{setensorsection}, we review the setup of the probe D7-brane in the AdS-Schwarzschild background, describing a SYM theory with finite $U(1)_B$ chemical potential and background electric and magnetic fields, discuss the polarization tensor of the flavor fields, and then compute the stress-energy tensor for the D7-brane. In section \ref{lossratessection} we use our result for the stress-energy tensor to calculate the energy and momentum loss rates, finding perfect agreement with field theory expectations based on hydrodynamics. Unfortunately, these loss rates also prevent us from obtaining new transport coefficients, such as $\alpha_{ij}$. In section \ref{zeroTsection} we consider a few special cases in which the loss rates vanish. In particular, we show that the zero-temperature limit of our result agrees with field theory expectations, giving a nice check of our method. In section \ref{conclusion} we conclude with some discussion and suggestions for future research. We collect some technical results in two appendices.

\section{Conserved Charges in AdS/CFT}
\label{chargessection}

\subsection{The Hamiltonian Framework}

Witten, in his seminal paper on the basics of the AdS/CFT correspondence \cite{Witten:1998qj}, provided two alternative points of view on how to relate field theory and bulk quantities. The most commonly used dictionary is the operator/field
mapping, where  every operator of the field theory is associated to a field in the bulk supergravity theory and the source strength and vacuum expectation value of the operator can be read off from the leading and sub-leading near-boundary behavior of the bulk field.

The alternative point of view is the Hamiltonian framework. If the bulk and boundary theories are truly equivalent as physical theories, they must have the same Hilbert space, and therefore every state in the field theory must map to a state of the bulk theory. As a consequence, for any given state, the expectation values of conserved charges such as
energy and momentum must agree between the bulk theory and the boundary theory.

This point of view is particularly useful when studying the correspondence in the presence of additional sources in the bulk, such as probe D-branes or semi-classical Nambu-Goto strings describing single quarks (or quark/anti-quark pairs). In these cases, using the standard operator/field dictionary to calculate the total energy requires solving for the
back-reaction of the probe, so that the stress-energy tensor of the field theory, $T_{\mu \nu}$, can then be read off from the asymptotic behavior of the bulk metric \cite{deHaro:2000xn}. Alternatively, the Hamiltonian framework guarantees that we can simply calculate the total energy of the bulk source and equate it with the total energy of the boundary solution.

This strategy has been used successfully for the study of Nambu-Goto strings, in particular for spinning strings, following the work of ref. \cite{Gubser:2002tv}. Here we will apply the same philosophy to probe D-branes. We will study spatially homogeneous solutions, so both bulk and boundary charges are simply the volume of space times a constant
density, and we can go one step further and directly equate the densities, rather than just the charges.

\subsection{Back-reaction}

At finite temperature, one subtlety that complicates the analysis is that the black hole horizon can carry part of the bulk charge. For example, for an electric test charge moving in the presence of a Reissner-Nordstrom black hole, the total boundary charge would be the sum of the test charge and the black hole charge. For sources that cross the horizon, as our probe D-branes will, the probe makes two contributions to the bulk energy: the energy of the part of the probe outside the horizon plus the change in mass of the black hole due to the presence of the probe. That is, for the purpose of calculating the energy of the probe, we cannot neglect the back-reaction of the probe.

Consider an action of the general form
\begin{equation}
S = N_c^2 S_{grav} + T_0 S_{probe},
\end{equation}
where we have extracted Newton's constant from the gravitational action and converted it to the field theory quantity $N_c^2$ and similarly we have extracted an overall normalization $T_0$ from the probe action, where $T_0 \ll N_c^2$ (in the case of a string $T_0$ is the string tension, which is order one in the large-$N_c$ counting, whereas for a probe D-brane $T_0$ is of order $N_c$). The back-reaction of the probe will only change the metric by a small correction of order $\ve=T_0/N_c^2$. Naively, we would therefore expect that the back-reaction is totally negligible to leading order in $\ve$. However, when calculating the energy density the order $\ve$ correction to the order $N_c^2$ energy density from the gravity part is of the same order, $T_0$, as the leading contribution from the probe action, so both must be included.

The only thermodynamic quantity that can be reliably calculated in the probe approximation is the free energy, which is simply minus the on-shell action. In this case, the order $\ve$ correction to the order $N_c^2$ on-shell action of the gravity sector is proportional to $\frac{\delta S_{grav}}{ \delta g_{\mu \nu}}$ evaluated on the background (uncorrected) metric, which vanishes by the equation of motion\footnote{One subtlety in this argument is the potential
contribution of boundary terms, which was analyzed in detail in ref. \cite{myersun}.
The variation of the Einstein-Hilbert term $\frac{1}{16 \pi G} \sqrt{g} R$
in the action vanishes in the bulk but leads to a finite boundary term of
the form
$\frac{1}{16 \pi G} \int_{\partial V} \sqrt{h} n^{\mu} v_{\mu}$ where $\partial
V$ is the boundary of spacetime, $n^{\mu}$ the outward-pointing normal, 
$h^{\mu \nu}$ the induced metric on the boundary, $h$ its determinant and
$$n^{\mu} v_{\mu} = n^{\mu} h^{\rho \sigma} (\nabla_{\rho} \, \delta g_{\mu 
\sigma} - \nabla_{\mu} \,  \delta g_{\rho \sigma}).$$ The second term
above cancels the variation of the Gibbons-Hawking term
$\int_{\partial V} \sqrt{h} K/(8 \pi G)$. In the standard
variational problem determining Einstein's equations one takes the
variation $\delta g_{\mu \nu}$ to vanish at the boundary and
so the first term which involves its derivative along the boundary
vanishes, too. 
In this case the variation of the action
vanishes identically and the variational problem is well-defined. 
In our case $\delta g_{\mu \nu}$ is the change in the metric
due to the backreaction of the brane, which in general will not vanish for
spacetime-filling branes that extend out to the boundary.
However, as this surviving
boundary term is proportional to a gradient of the backreaction
along the boundary it will vanish for field theory configurations \textit{without}
spatial gradients, as we consider in this work.}. 
The only contribution to the free energy at order $T_0$ is thus the probe's on-shell action (evaluated on the background (uncorrected) metric). This fact has been used in most recent thermodynamic studies of flavor physics using probe D-branes, for example refs. \cite{Mateos:2006nu,Albash:2006ew,Karch:2006bv,Mateos:2007vn}, but has been less appreciated in the study of string probes. We will illustrate this point in two examples below.

An important conclusion to draw for probe D-brane systems is that for quantities where the order $N_c^2$ term from the background vanishes (so that the contribution of the plasma dual to the bulk degrees of freedom is only of order one) the only contribution at order $N_c$ comes from the probe D-branes, and we can safely neglect the back-reaction. In particular, this allows us to calculate momentum currents in the static black hole background reliably from the probe D-brane stress-energy tensor. Energy densities, on the other hand, will receive contributions from both the probe and the horizon.

\subsection{Examples}
\label{examples}

\subsubsection{A Simple Toy Model}
\label{toymodel}

Let us first demonstrate the basic point in a simple exactly-solvable toy model. Consider pure Einstein gravity with a negative cosmological constant coupled to a spacetime-filling probe D-brane of tension $\frac{T_0}{16 \pi G}$; for a probe D-brane $T_0/G$ is a small parameter of order $N_f/N_c$. The Euclidean action is
\beq
S = \frac{1}{16 \pi G} \, \int d^{d+1}x \, \sqrt{g} \left ( R + \frac{d (d-1)}{L^2} - T_0 \right ).
\eeq
In the absence of the brane ($T_0=0$) the equations of motion that
follow from this action allow planar AdS-Schwarzschild solutions
\beq
\label{rcoords}
ds^2
=  h(r) \, dt^2 + \frac{dr^2}{h(r)} + \frac{r^2}{L^2} \, d \vec{x}^2 ,
\qquad h(r) = \frac{r^2}{L^2} - \frac{r_h^d}{L^2 r^{d-2}}.
\eeq
$t$ is the time coordinate and $d\vec{x}^2$ is the metric of three-dimensional Euclidean space, while $r$ is a radial coordinate that runs from the horizon at $r=r_h$ to the boundary at $r = \infty$. The Ricci tensor, Ricci scalar, and temperature associated with this geometry are
\beq
R_{\mu \nu} = - \frac{d}{L^2} \, g_{\mu \nu}, \,\,\,\,\,\,\,\,\,\, R = -
\frac{d (d+1)}{L^2}, \,\,\,\,\, T = \frac{1}{\beta} = \frac{r_h}{L^2}
\frac{d}{4 \pi},
\eeq
so that $L$ is the radius of curvature of the space\footnote{This is the only subsection of the paper in which we write the $L$ explicitly. In all other parts of the paper we will use units in which $L\equiv1$. Also, this subsection and the next are the only places in the paper where we use a Euclidean-signature metric. In all other subsections we use a Lorentzian-signature metric.}.

We identify the free energy density of the field theory as minus the on-shell action (divided by $\beta$ and the volume $\int d\vec{x}$), which is\footnote{Here we have added counterterms to cancel an $r \ra \infty$ divergence of the on-shell action (interpreted as a UV divergence in the field theory). The counterterms produce a finite contribution that is $-\frac{1}{2}$ times the contribution from the $r = r_h$ boundary of the integral, so that the overall answer is $1/2$ times the $r=r_h$ term. In ref. \cite{Witten:1998zw}, where background subtraction was used instead of counterterms, the factor of $1/2$ arises from making sure that both black hole and the thermal AdS background subtraction are really at the same temperature.}
\beq
 f = - \frac{1}{16 \pi G L^{d+1}} \, r_h^{d} = - \frac{1}{16 \pi G}\, \left (
\frac{4 \pi}{d} \right)^d \, L^{d-1} T^d.
\eeq
The entropy density $s$, energy density $\epsilon$, and pressure $P$ can be obtained from $f$ via standard thermodynamic relations or directly from the gravity side: the entropy is the Bekenstein-Hawking entropy associated with the horizon area and the stress-energy tensor was calculated in ref. \cite{Balasubramanian:1999re}. The results are
\beq P = -f, \,\,\,\,\,\,\,\,\,\, \epsilon = (d-1) P,
\,\,\,\,\,\,\,\,\,\, s T = d \, P
\eeq
as appropriate for a conformal field theory.

The effects of the tension $T_0$ can easily be incorporated analytically, as $T_0$ simply shifts the cosmological constant. All thermodynamic quantities only change by replacing the old curvature radius $L$ with the new curvature $l$ subject to the relation
\beq
\frac{d (d-1)}{L^2} - T_0 = \frac{d (d-1)}{l^2}
\eeq
or, expanding to leading order in $t= T_0 L^2$,
\beq
l = L \left ( 1 + \frac{t}{2 d (d-1)} \right ) .
\eeq
Correspondingly, the shift in free energy is
\beq
\delta f = \frac{t}{2 d} f .
\eeq
From $\delta f$ we can also calculate $\delta \epsilon = - (d-1) \delta f$ and $T \delta s = - d \delta f$.

We want to confirm that the effect of $T_0$ can also be correctly captured directly by a probe calculation. As is well-documented in the literature, the probe D-brane's contribution to the free energy should simply be given by $- \beta^{-1}$ times the on-shell action of the D-brane in the original, uncorrected AdS black hole background. Naively, the gravitational on-shell action evaluated on the back-reacted solution (that is, the solution incorporating the order $t$ effects due to the D-brane) should also contribute at order $t$, but this contribution is proportional to the variation of the action with respect to the metric evaluated on the zeroth-order solution, which vanishes because the zeroth-order solution extremizes the gravitational action. Indeed, we find that the D-brane's on-shell action gives
\beq
\delta f_{probe} = - \frac{T_0}{32 \pi G L^{d-1}} \frac{r_h^d}{d} = \frac{t}{2d} f
\eeq
in perfect agreement with the analytic answer expanded to order $t$. Calculating the D-brane's energy from its stress-energy tensor, which we will denote $U_{\mu \nu}$, we find
\beq
\delta \epsilon_{probe} = -\int dr \, \sqrt{g} \,\, U^t_{~t} = -\frac{T_0}{16\pi G}  \int dr \sqrt{g}=  \delta f.
\eeq
The energy density is equal to minus the on-shell action, instead of the expected $(d-1)$ times the on-shell action.

Note what is missing: we haven't accounted for the part of the probe D-brane hidden behind the horizon. The only effect of that part of the D-brane is to increase the horizon area. The total energy in the bulk has to include this change in black hole mass to count the total energy in the bulk correctly. Indeed, the first law of black hole thermodynamics, which is purely a statement of classical gravity, tells us that the change in horizon area comes with a change in black hole mass given by
\beq
\delta \epsilon_{bh} = T \delta s_{bh} = -d \delta f
\eeq
so that indeed
\beq
\delta \epsilon_{bh} + \delta \epsilon_{probe} = -(d-1) \delta f.
\eeq
The contribution of the D-brane alone only accounts for a small fraction of the total bulk energy, and moreover even contributes with the ``wrong'' sign. We will encounter this issue in various forms throughout this paper.

\subsubsection{Finite-Temperature Correction to Quark Mass}

Our second example, more relevant to finite-temperature calculations performed in the AdS dual of strongly-coupled ${\cal N}=4$ SYM, is the mass shift of a heavy quark in the thermal ${\cal N}=4$ SYM plasma, which was first considered in ref. \cite{Herzog:2006gh}. As we will explain below, dynamical flavors in the SYM theory are represented by probe D-branes in AdS that extend down to some radial position $r_m$ (in the coordinates of eq. (\ref{rcoords})). In the gravitational theory, a single quark is represented by a string stretching all the way from the horizon to the probe D-brane. At zero temperature the energy of such a string is given by $m=\frac{1}{2 \pi \alpha'} \, r_m$. For this static solution, $m$ is identified as the mass of the quark. As we argued above, the reliable quantity to evaluate is the free energy, \textit{i.e.} minus the on-shell action. For the string worldsheet under consideration, $g_{rr} g_{tt}=1$, so even at finite temperature we find that the on-shell action is the length of the string times the circumference of the Euclidean time circle; the only difference at finite $T$ is that the string only stretches from the horizon at $r_h$ to $r_m$. The result is that we find for the free energy of the quark\footnote{The expression below is the answer for a quark of fixed color. For a quark of unspecified color $F$ receives an additional trivial contribution of $\log(N_c)$.}
\beq
F = \frac{1}{2 \pi \alpha'} \left( r_m - r_h \right) =  m - \frac{1}{2} \sqrt{\lambda} \, T
\eeq
leading to a temperature-independent entropy  of $S=\frac{1}{2} \sqrt{\lambda}$. Using thermodynamics, we can compute the energy of the quark at finite temperature,
\begin{equation}
E = F + T S =m.
\end{equation}
That is, the energy of the quark is independent of temperature.

In ref. \cite{Herzog:2006gh} the energy of the stretched string was calculated directly via the canonical momenta of the Nambu-Goto action, with the result $E=m - \frac{1}{2} \sqrt{\lambda} T$. This is clearly not the correct answer for the energy of the string, as it contradicts basic thermodynamics. Apparently the back-reaction of the string on the black hole will have to change the mass of the black hole by $\frac{1}{2} \sqrt{\lambda}$ to account for the correct energy $E$ of the quark. In ref. \cite{Herzog:2006gh}, $\frac{1}{2} \sqrt{\lambda}$ was somewhat loosely referred to as the change in mass of the quark. A precise statement of what this quantity really represents is the change in free energy of the quark.

Let us summarize the important lessons of this section. In both of our examples, the probe's stress-energy tensor (or, for the string probe, canonical momenta) gave the \textit{free energy} precisely, \textit{not} the energy. In other words, the probe's stress-energy tensor completely misses the entropy contribution. This is to be expected, as the contribution to the entropy takes the form of an expanding horizon, which is a result of the back-reaction of the probe. The free energy is given by the probe action, the entropy is given by the back-reaction, and the energy is given by both the probe action and the back-reaction.

\section{The Stress-Energy Tensor of Flavor Fields}
\label{setensorsection}

\subsection{The D7-brane Solution}
\label{d7solution}

In this subsection we present the supergravity solution describing massive hypermultiplets propagating through an $\N=4$ SYM plasma with finite $U(1)_B$ chemical potential and in the presence of external electric and magnetic fields. We will also provide a rough sketch of the SYM theory phase diagram, to give a sense of where the results of later sections will be valid. Most of this subsection is a summary of refs. \cite{Karch:2007pd,O'Bannon:2007in}. For readers already familiar with refs. \cite{Karch:2007pd,O'Bannon:2007in}, we recommend skipping to the next subsection and consulting this subsection as needed.

We use a Fefferman-Graham \cite{Fefferman} form for the $AdS_5$ metric
\beq
ds^2_{AdS_5}  = \frac{dz^2}{z^2} + \, g_{tt} \, dt^2 + g_{xx} \, d\vec{x}^2
\eeq
where $z$ is the $AdS_5$ radial coordinate. The boundary is at $z = 0$. When we need an explicit form of the metric, we will use
\beq
ds_{AdS_5}^2 = \frac{dz^2}{z^2} - \frac{1}{z^2} \frac{(1 - z^4 / z_H^4)^2}{1+z^4/ z_H^4} \, dt^2 + \frac{1}{z^2} (1+z^4 / z_H^4) \, d\vec{x}^2.
\eeq
The black hole horizon is at $z = z_H$ with $z_H^{-1} = \frac{\p}{\sqrt{2}} \, T$. Here we are using units in which the radius of $AdS_5$ is equal to one. In these units, we convert from string theory to SYM quantities using $\alpha'^{-2} = \lambda$.

We will use an $S^5$ metric of the form
\beq
ds^2_{S^5} = d\theta^2 + \sin^2 \theta \, ds^2_{S^1} + \cos^2 \theta \, ds^2_{S^3}
\eeq
where $\theta$ is an angle between zero and $\pi/2$ and $ds^2_{S^1}$ and $ds^2_{S^3}$ are metrics for a unit-radius circle and 3-sphere, respectively.

We next introduce $N_f$ probe D7-branes. The relevant part of their action will be the Born-Infeld term,
\beq
S_{D7} = - N_f T_{D7} \int d^8 \zeta \sqrt{-det\left(g_{ab} + (2\pi\alpha') F_{ab} \right)}.
\eeq
Here $T_{D7}$ is the D7-brane tension, $\zeta^a$ are the worldvolume coordinates, $g_{ab}$ is the induced worldvolume metric, and $F_{ab}$ is the $U(1)$ worldvolume field strength. The D7-branes will be extended along all of the $AdS_5$ directions, as well as the $S^3$ directions inside the $S^5$.

Our ansatz for the worldvolume fields will include the worldvolume scalar $\theta(z)$. The D7-brane induced metric is then identical to the background metric, except for the radial component, which is $g_{zz} = \frac{1}{z^2} + \theta'(z)^2$, where prime denotes differentiation with respect to $z$. The supergravity field $\theta(z)$ is dual to the SYM theory operator which, roughly speaking, is the mass operator of the flavor fields. We will denote it as $\Om$. The precise form of $\Om$ appears for example in ref. \cite{Kobayashi:2006sb}; just thinking of $\Om$ as the mass operator will be sufficient for our purposes. We can identify the mass and expectation value $\Omv$ from $\theta(z)$'s asymptotic form,
\beq
\theta(z) = c_1 \, z + c_3 \, z^3 + O(z^5).
\eeq
The mass $m$ of the flavor fields is then $m = \frac{c_1}{2\pi\alpha'}$, and the expectation value is given by $\Omv \propto - 2 c_3 + \frac{1}{3} c_1^3$ \cite{Karch:2005ms,Karch:2006bv}.

We discuss $\theta(z)$'s equation of motion and boundary conditions below, around eq. (\ref{lteffaction}). We will soon discuss other D7-brane worldvolume fields, in particular worldvolume electric and magnetic fields; $\theta(z)$'s equation of motion will of course depend on their values. We can in principle solve for $\theta(z)$ numerically, but in what follows we will not do so. Instead we will write our results in terms of $\theta(z)$ and make use of special cases where we know exact solutions. For massless flavors ($c_1=0$), $\theta(z)=0$ is always an exact solution for any values of the external fields, and corresponds to massless quarks with $\Omv = 0$. Our D7-brane will have a worldvolume magnetic field, however, so we must remember that, as shown in refs. \cite{Filev:2007gb,Filev:2007qu,Albash:2007bk,Erdmenger:2007bn}, in the presence of a background magnetic field $\theta=0$ is typically only meta-stable, and the true ground state has a non-trivial $\theta(z)$ even when $c_1=0$. In the field theory we interpret this as spontaneous chiral symmetry breaking induced by the magnetic field. Still, we will use the meta-stable $\theta=0$ solution as a simple example to confirm that our bulk expressions agree with field theory expectations.

In the AdS-Schwarzschild background, two topologically distinct types of D7-brane embeddings exist, the so-called ``Minkowski'' and ``black hole'' embeddings. In a Minkowski embedding, $\theta(z)$ describes a D7-brane for which the $S^3$ inside the $S^5$ ``slips'' as the D7-brane extends from the boundary into the bulk of $AdS_5$, eventually collapsing to zero volume, \textit{i.e.} $\theta(z)$ begins at zero at the boundary and reaches $\frac{\pi}{2}$ at some $z > z_H$, so that the D7-brane smoothly terminates outside the horizon. In a black hole embedding, $\theta(z)$ never reaches $\frac{\pi}{2}$ outside the horizon, so the D7-brane intersects the horizon. In the absence of worldvolume electric and magnetic fields, a first-order phase transition from Minkowski to black hole embeddings occurs as $m/T$ is lowered through a critical value \cite{Babington:2003vm,Kirsch:2004km,Mateos:2007vn,Mateos:2006nu,Albash:2006ew,Karch:2006bv}. The transition is often called a ``meson melting'' transition: Minkowski embeddings give rise to a gapped, discrete meson spectrum in the SYM theory, while black hole embeddings give rise to a gapless, continuous meson spectrum \cite{Hoyos:2006gb}.

The $U(N_f)$ gauge invariance of the coincident D7-branes is dual to the $U(N_f)$ symmetry of the mass-degenerate flavor fields in the SYM theory. We identify the $U(1)$ subgroup as baryon number, $U(1)_B$. The D7-brane worldvolume Abelian gauge field $A_{\mu}$ is dual to the SYM $U(1)_B$ current $J^{\mu}$, so to introduce a finite $U(1)_B$ density in the SYM theory, we must introduce the worldvolume gauge field $A_t(z)$. We may identify the $U(1)_B$ chemical potential $\mu$ and the density $\jt$ from the asymptotic form\footnote{A good question is why $\mu$ is physical: it is constant, so normally we could gauge it away. We discuss the boundary condition on $A_t(z)$, which forbids such gauge transformations, as well as the boundary conditions on the other gauge fields, in appendix \ref{boundaryconditions}.} of $A_t(z)$:
\beq
A_t(z) = \mu + c \, z^2 + O\left(z^4\right)
\eeq
where the constant $c$ gives the density: $\jt \propto c$ \cite{Kobayashi:2006sb}.

As shown in ref. \cite{Kobayashi:2006sb}, with nonzero $A_t(z)$, only black hole embeddings are allowed. The physical reason is simple: with nonzero $A_t(z)$ the D7-brane has an electric field pointing in the radial direction, $F_{zt}$. These field lines are sourced by a density of strings. The force that such a density of strings exerts on the D7-brane overcomes the tension of the D7-brane, pulling the D7-brane into the horizon. The strings are then ``hidden'' behind the horizon, where the field lines may safely terminate. Numerical \cite{Kobayashi:2006sb} and analytic \cite{Karch:2007br} analysis of D7-brane embeddings has confirmed that with nontrivial $A_t(z)$ the only physical D7-brane embeddings are black hole embeddings. When we work with nonzero $\jt$ in the SYM theory, our D7-brane will always intersect the horizon.

The first-order ``meson melting'' phase transition discovered at zero density has been shown to persist to finite density, producing a line of first-order transitions that ends in a critical point \cite{Kobayashi:2006sb}. A region of the phase diagram near the line of transitions is also known to be thermodynamically unstable \cite{Kobayashi:2006sb}. Our results will not apply in the unstable region.

To introduce perpendicular electric and magnetic fields, and the resulting currents $\jx$ and $\jy$, we also include in our ansatz the gauge field components \cite{Karch:2007pd,O'Bannon:2007in}
\beq
A_x(t,z) = -Et + h_x(z), \qquad A_y(x,z) = Bx + h_y(z).
\eeq
In each case, the leading term is a non-normalizable mode that introduces an external field into the SYM theory. Choosing a gauge in which $A_z = 0$, we can write the nonzero elements of $F_{ab}$ as:
\beq
F_{tx} = - E, \qquad F_{xy} = B
\eeq
\beq
F_{zt} = A'_t, \qquad F_{zx} = A'_x, \qquad F_{zy} = A'_y.
\eeq

To date, the SYM theory in the presence of external fields has been analyzed (via AdS/CFT) with only $E$ \cite{Erdmenger:2007bn,Albash:2007bq} or $B$ \cite{Filev:2007gb,Filev:2007qu,Albash:2007bk,Erdmenger:2007bn} alone, not both simultaneously, and only with zero $U(1)_B$ density. As mentioned above, at zero temperature the magnetic field triggers spontaneous chiral symmetry breaking and also produces a Zeeman-like effect on the meson spectrum. The magnetic field also increases the meson melting transition temperature. Heuristically, the magnetic field ``holds mesons together,'' and exhibits a kind of ``magnetic catalysis'' of chiral symmetry breaking. Indeed, in supergravity language, for sufficiently large magnetic field only Minkowski embeddings exist. On the other hand, an electric field lowers the meson melting temperature, which is easy to understand intuitively: the electric field will pull quarks and anti-quarks (or squarks and anti-squarks) in opposite directions, so that we do not need to heat up the system as much to ``melt'' mesons.

Our ansatz only involves gauge field components in the four-dimensional $(z,t, x, y)$ subspace, so the D7-brane action takes the form of the four-dimensional Born-Infeld action, times some ``extra'' factors coming from the extra dimensions,
\beq
S_{D7} = - \N \int dz \cos^3\th \, g_{xx}^{1/2} \, \sqrt{- g - (2 \p \a')^2 \frac{1}{2} \, g \, F^2 - (2 \p \a')^4 \frac{1}{4} \left(  F \wedge F \right)^2}.
\label{originaldbi}
\eeq
We have divided both sides of eq. (\ref{originaldbi}) by the volume of ${\mathbb{R}^{3,1}}$, so that technically $S_{D7}$ is an action density. We have defined $g=g_{zz} \, g_{tt} \, g_{xx}^2$ as the determinant of the induced metric in the $(z,t,x,y)$ subspace. We have also introduced the constant $\N$, which, using $T_{D7} = \frac{\alpha'^{-4} g_s^{-1}}{(2\pi)^7} = \frac{\lambda N_c}{2^5 \pi^6}$, is
\beq
\N \equiv N_f T_{D7} V_{S^3} = \frac{\lambda}{(2\pi)^4} N_f N_c
\eeq
where $V_{S^3} = 2 \pi^2$ is the volume of a unit-radius $S^3$. Writing $F^2 = F^{\m\n} F_{\m\n}$, where Greek indices run over $(z,t,x,y)$, we have explicitly

\begin{subequations}
\label{gaugeadef}
\beq
\frac{1}{2} \, g \, F^2 = g_{xx}^2 A_t'^2 + g_{tt} g_{xx} A_x'^2 + g_{tt} g_{xx} A_y'^2 + g_{zz} g_{xx} E^2 + g_{zz} g_{tt} B^2
\label{a2def}
\eeq
\beq
\frac{1}{4} \left( F \wedge F\right)^2 = B^2 A_t'^2 + E^2 A_y'^2 - 2 E B A_t' A_y'.
\label{a4def}
\eeq
\end{subequations}
Starting now, $L$ will denote the Lagrangian density, albeit with an unconventional sign choice, introduced for future convenience: $S_{D7} = - \int dz \, L$.

Clearly the action only depends on the $z$ derivatives of $A_t$, $A_x$ and $A_y$, so classically the system has three constants of motion. As shown in refs. \cite{Karch:2007pd,O'Bannon:2007in}, we can identify these as the components of the $U(1)_B$ current density in the SYM theory,
\beq
\label{currentdefinition}
\langle J^i \rangle = \frac{\delta L}{\delta A'_i}.
\eeq
Our system thus has a nonzero $U(1)_B$ density $\langle J^t \rangle$ as well as $U(1)_B$ currents $\jx$ and $\jy$. Given these constants of motion, we can solve algebraically for the derivatives of the gauge field (the field strength components):
\beq
\label{ataxsol}
A_t'(z) = - \frac{\sqrt{g_{zz} |g_{tt}|}}{g_{xx}} \frac{\jt \xi - B a}{\sqrt{\xi \chi - a^2}}, \qquad A_x'(z) = \sqrt{\frac{g_{zz}}{|g_{tt}|}} \frac{\jx \xi}{\sqrt{\xi \chi - a^2}}
\eeq
\beq
\label{aysol}
A_y'(z) =  \sqrt{\frac{g_{zz}}{|g_{tt}|}} \frac{\jy \xi + E a}{\sqrt{\xi \chi - a^2}}
\eeq
where
\beq
\label{xichiadef}
 \xi =  |g_{tt}| g^2_{xx} + (2 \pi \alpha')^2
(|g_{tt}| B^2 - g_{xx} E^2 ), \qquad
a =(2 \pi \alpha')^2 ( |g_{tt}| \langle J_t \rangle B + g_{xx} \langle J_y \rangle E )
\eeq
\beq
\chi = |g_{tt}| g^3_{xx} {\cal N}^2 (2 \pi \alpha')^4 \cos^6\theta + (2 \pi \alpha')^2 (|g_{tt}| \langle J_t \rangle^2 - g_{xx} ( \langle J_x \rangle^2 + \langle J_y \rangle^2 )).
\eeq

In terms of these quantities the on-shell action, which can be identified as minus the free energy density in the grand canonical ensemble, is given by
\beq
S_{D7} = - \N^2 (2\pi\alpha')^2 \int dz \, \cos^6\th \, g_{xx}^2 \sqrt{g_{zz} |g_{tt}|} \, \frac{\xi}{\sqrt{\xi \chi - a^2}}.
\label{effaction}
\eeq
The equation of motion for the worldvolume scalar $\theta(z)$ can be derived in two ways, either from the Legendre-transform of the action, which we denote as $\hat{S}_{D7}$,
\bea
\label{lteffaction}
\hat{S}_{D7} & = & S_{D7} - \int dz \left ( F_{zt} \frac{\d S_{D7}}{\d F_{zt}} + F_{zx} \frac{\d S_{D7}}{\d F_{zx}} + F_{zy} \frac{\d S_{D7}}{\d F_{zy}} \right ) \\ & = & - \frac{1}{(2\p\a')^2} \int dz \, g_{zz}^{1/2} |g_{tt}|^{-1/2} g_{xx}^{-1} \sqrt{\xi \chi - a^2}, \nonumber
\eea
or by varying eq. (\ref{originaldbi}) and then plugging in the solutions in eq. (\ref{ataxsol}) and (\ref{aysol}).

To complete the D7-brane solution, we must specify boundary conditions on $\theta(z)$. For Minkowski embeddings, we have $\theta(z_m) = \frac{\pi}{2}$ for some $0\leq z_m \leq z_H$ and $\theta'(z_m) = \infty$ to avoid a conical singularity \cite{Karch:2006bv}. For black hole embeddings, the boundary conditions are $\theta'(z_H)=0$ for the embedding to be static, while $\theta(z_H) \in \left[0,\frac{\pi}{2} \right]$ is a free parameter. Notice that each case has a free parameter (the value of $z_m$ or $\theta(z_H)$) which maps in a one-to-one fashion to the asymptotic parameter $c_1 \propto m$. For example, with nonzero $A_t(z)$ (hence only black hole embeddings), the limits are $\theta(z_H) = 0$ maps to $c_1=0$ (when $B$ is nonzero, this is the meta-stable massless solution) and $\theta(z_H) \rightarrow \frac{\pi}{2}$ maps to $c_1 \rightarrow \infty$.

\subsection{The Polarization Tensor}
\label{polarizationsection}

In the next subsection we will use the D7-brane solution above to compute the stress-energy tensor of the flavor fields. Some contributions to the stress-energy tensor come just from the electric polarization and the magnetization of the medium, however, as we will now explain.

Even in an equilibrium system, background electric and magnetic fields produce non-vanishing momentum currents due to polarization effects. As reviewed for example in ref. \cite{Hartnoll:2007ih} (and references therein), even in equilibrium we expect a contribution to $\langle T_{\mu \nu} \rangle$ of the form
\beq
\langle T^{\mu}_{~\nu} \rangle_{pol} = M^{\mu}_{~ \sigma} \, F^{\sigma}_{~ \nu}.
\eeq
where $M^{\mu \nu}$ is the polarization tensor,
\beq
M^{\mu \sigma} = - \frac{\delta \Omega}{\delta F_{\mu \sigma}},
\eeq
with $\Omega$ the free energy density (and where we take the derivative with other variables held fixed). The components of $M^{\mu \sigma}$ with one $t$ index and one spatial index are electric polarizations while components with two spatial indices are magnetizations. The full energy-momentum tensor $\langle T^{\mu}_{~\nu} \rangle$ then divides into two pieces:
\beq
\langle T^{\mu}_{~\nu} \rangle = \langle T^{\mu}_{~\nu} \rangle_{fluid} + \langle T^{\mu}_{~\nu} \rangle_{pol}
\eeq
where, for example, $\langle T^{t}_{~i} \rangle_{fluid}$ corresponds to the genuine momentum current due to the flow in the medium. As reviewed in ref. \cite{Hartnoll:2007ih}, both $\langle T^{\mu}_{~\nu} \rangle$ and $\langle T^{\m}_{~\n} \rangle_{fluid}$ obey the same conservation equation,
\beq
\partial^{\mu} \langle T_{\mu \nu} \rangle = F_{\nu \rho} \langle J^{\rho} \rangle,
\eeq
but only $\langle T^{\m}_{~\n} \rangle_{fluid}$ represents observable quantities that can couple to external probes of the system, and hence is the appropriate object to use when studying transport, for example, when computing transport coefficients. In particular, we should use $\langle T^t_{~i} \rangle_{fluid}$ to identify the heat current densities (that we discussed in the introduction),
\beq
\langle Q_i \rangle \equiv \langle T^t_{~i} \rangle_{fluid} - \mu \langle J_i \rangle.
\eeq

In gauge-gravity duality, we identify $\Omega = - S_{D7}$, where here $S_{D7}$ is the D7-brane action evaluated on a particular solution for the worldvolume fields. For our choice of background electric and magnetic fields, the $x$ and $y$ components of the polarization, $M^{ti}$ ($i=x,y$), and the magnetization $M^{xy}$ will be non-vanishing:
\beq
M^{ti} = - \frac{d S_{D7}}{d E_i}, \qquad M^{xy} =  \frac{d S_{D7}}{d B}.
\eeq
Notice that both the electric and magnetic fields must be nonzero for the polarizations to contribute to the momentum densities:
\beq
\langle T^t_{~x} \rangle = \langle T^t_{~x} \rangle_{fluid} - M^{ty} B, \qquad \langle T^t_{~y} \rangle = \langle T^t_{~y} \rangle_{fluid} + M^{tx} B.
\eeq

Notice also that $M^{ty}$ will be nonzero even though we have not introduced a background electric field in the $y$ direction. Suppose we did introduce an electric field in the $y$ direction, $E_y$. We then easily find that, due to the $F \wedge F$ term in the action, eq. (\ref{originaldbi}), taking the derivative with respect to $E_y$ and then setting $E_y = 0$ produces a nonzero result. Indeed, a little algebra shows that
\beq
\frac{d S_{D7}}{d E_y} = \frac{1}{E} \int dz \left( A_x' \jy - A_y' \jx \right),
\eeq
which we will use to simplify expressions in the sequel. A useful fact to remember is that $M^{ty}=0$ when $\jt = 0$: when $\jt=0$ we can use the result for the conductivity in appendix \ref{conductivity} to show that $\jy=0$, and from the explicit solution for $A_y'$ in eq. (\ref{aysol}) we can also show that $\jt=0$ implies $A_y'=0$.

The calculation of $M^{tx}$ and $M^{xy}$ from $S_{D7}$ is more complicated. Consider for example $M^{tx}$. In field theory terms, we need to compute $\left . \frac{ d \Omega} { d E} \right |_{T, \mu, B}$. We start with eq. (\ref{originaldbi}), evaluated on a particular solution. The on-shell action $S_{D7} = - \Omega$ will then have explicit $E$ dependence, as well as implicit dependence through the solutions for $\theta(z)$ and the worldvolume gauge fields. We thus employ the chain rule\footnote{We are using arguments similar to those in refs. \cite{Mateos:2007vn,Albash:2007bk,O'Bannon:2008bz}.},
\beq
\frac{d S_{D7}}{d E} = - \int dz \, \left [ \frac{\partial L}{\partial E} \, + \, \frac{\partial \theta}{\partial E} \frac{\partial L}{\partial \theta} \, + \, \frac{\partial \theta'}{\partial E} \frac{\partial L}{\partial \theta'} \, + \, \sum_{i=t,x,y} \frac{\partial A_i'}{\partial E} \frac{\partial L}{\partial A_i'} \right].
\eeq
Notice in paricular that in the $\frac{\partial L}{\partial E}$ term the derivative only acts on the explicit $E$ dependence in $L$ (\textit{i.e.} on the explicit factors of $E$ appearing in eqs. (\ref{a2def}) and (\ref{a4def})). We then use the fact that partial derivatives commute to write $\frac{\partial}{\partial E} \frac{\partial}{\partial z} = \frac{\partial}{\partial z} \frac{\partial}{\partial E}$, and integrate by parts to find
\bea
\label{bulkpolarizationequation}
\frac{d S_{D7}}{d E} & = & - \int dz \, \left [ \frac{\partial L}{\partial E} \, + \, \left( \frac{\partial L}{\partial \theta} \, - \, \frac{\partial}{\partial z} \frac{\partial L}{\partial \theta'} \right) \frac{\partial \theta}{\partial E} \, - \, \sum_{i=t,x,y} \frac{\partial A_i}{\partial E} \frac{\partial}{\partial z} \frac{\partial L}{\partial A_i'} \right] \nonumber \\ & & \,\,\,\,\,\,\, - \left . \frac{\partial \theta}{\partial E} \frac{\partial L}{\partial \theta'} \right |_{0}^{z_H} - \left . \sum_{i=t,x,y} \frac{\partial A_i}{\partial E} \frac{\partial L}{\partial A_i'} \right |_{0}^{z_H}.
\eea
Obviously, of the terms under the integral, the term in parentheses and the terms in the sum over $i$ vanish due to the equations of motion. That leaves the $\frac{\partial L}{\partial E}$ term under the integral, and the boundary terms. We will not compute $\frac{\partial L}{\partial E}$ in what follows, nor will we discuss the boundary terms. Our main point is that the only contribution to the polarization from the bulk of $AdS_5$ comes from $\frac{\partial L}{\partial E}$. A similar statement applies for the magnetization $M^{xy}$, \textit{i.e.} the only bulk term comes from $\frac{\partial L}{\partial B}$. In sections \ref{BGE} and \ref{EGGB} we will take limits in which the boundary terms vanish (or are negligibly small), in which case the bulk terms become the only contributions to the polarization tensor.

In the next subsection we will see factors of $\frac{\partial L}{\partial E}$, $\frac{\partial L}{\partial E_y}$, and $\frac{\partial L}{\partial B}$ appearing in the stress-energy tensor. Most of these arise from the expected contribution to $\langle T_{\mu \nu} \rangle$ from $\langle T^{\mu}_{~\nu} \rangle_{pol}$.

\subsection{ The Stress-Energy Tensor}
\label{d7stressenergytensor}

Our goal now is to compute the contribution that the flavor fields make to the stress-energy tensor of the SYM theory, using the above holographic setup. We have chosen to work in the Hamiltonian framework. If $p_i$ denotes the momentum associated with the flavor fields in the SYM theory, with $i = x,y$ our momenta of interest, then in the Hamiltonian framework we identify the conserved charges
\beq
p_i = \int dt \, d\vec{x} ~\Ti = \int dt \, d\vec{x} \,dz \, d^3\alpha ~\sqrt{-g_{D7}} \, \, \Ui.
\eeq
The $\alpha$ are coordinates on the $S^3$ wrapped by the D7-branes and $g_{D7}$ is the determinant of the
induced metric on the D7-branes. We have introduced the notation $\Ui$ as the D7-branes' momentum density, reserving the notation $\Ti$ for the expectation value of the flavor fields' momentum density in the SYM theory. Note that in order to form a covariant quantity, we must include a factor of $\sqrt{-g_{D7}}$ in the integral.  In principle, a similar factor should also appear in the four-dimensional integral, but as the SYM theory lives in flat space, this factor is unity. Furthermore, if the energy-momentum tensors are independent of the four spacetime coordinates, then the integrals over $dt \, d\vec{x}$ will only produce a factor of the spacetime volume, so that we can equate the momentum densities directly:
\beq
\label{FTT}
\Ti = \int dz \, d^3\alpha ~\sqrt{-g_{D7}} \, \, \Ui.
\eeq

Our task is thus to compute the stress-energy tensor of the D7-branes. Let us first define notation. Let
\beq
\label{d7setensor}
\Theta^a_{~b} \equiv \int \, dz \, d^3\alpha ~\sqrt{-g_{D7}} \, U^{a}_{~b}.
\eeq
When the indices $a$ and $b$ are in SYM theory directions, we can identify $\langle T^a_{~b}\rangle = \Theta^a_{~b}$. The indices $a$ and $b$ can also be in the $z$ or $S^3$ directions, however, in which case the SYM theory interpretation requires more effort. In what follows we will be able to translate some, but not all, such components into SYM theory quantities.

We can compute $\Theta^a_{~b}$ in two different ways. We can of course directly compute the variation of the D7-brane action, $S_{D7}$, with respect to the background metric. Alternatively, because the momenta are the generators of translation symmetries, we can derive the tensor components via a Noether procedure. The two methods must agree up to boundary terms\footnote{See for example sections 7.3 and 7.4  of ref. \cite{Weinberg:1995mt}.}. Indeed, we have used both methods and have found perfect agreement (not just agreement up to boundary terms). As the calculation by variation of the action is somewhat lengthy, we include it in appendix \ref{d7stressvariation}. The result of the Noether procedure is
\beq
\label{noetherstressenergy}
\Theta^a_{~b} = -\int dz \, \left( L \, \delta^{a}_{~b} + 2 F_{c b} \frac{ \delta L }{\delta F_{a c} } - \partial_{b}\theta \, \frac{\delta L}{\delta \partial_a \theta} \right),
\eeq
where we have performed the trivial integration over the $S^3$, since our ansatz for the worldvolume fields is independent of these directions.

Note that the bulk theory does not actually have translation invariance in the $z$ direction. Calculating $T^{\m}_{~z}$ using the Noether procedure may thus seem suspicious. As mentioned above, however, even for these components we have verified explicitly that the Noether calculation agrees with the calculation via variation of the background metric.

We expect the last term in eq. (\ref{noetherstressenergy}) to contribute to $T^{z}_{~z}$, given our ansatz $\theta(z)$. We find, however, that the last term in eq. (\ref{noetherstressenergy}) also contributes to the $T^{i}_{~z}$ components with $i = t,x,y$. To see why, suppose we allow $\theta$ to depend on $t,x,y$. We then find that, due to the $g F^2$ term in eq. (\ref{originaldbi}), taking the derivatives $\frac{\delta L}{\delta \partial_t \theta}$, $\frac{\delta L}{\delta \partial_x \theta}$, and $\frac{\delta L}{\delta \partial_y \theta}$ and then setting $\partial_t \theta = \partial_x \theta = \partial_y \theta = 0$ produces a nonzero result. This is very similar to what we saw in the last subsection, where the $y$ polarization was nonzero even though our ansatz has $E_y=0$. We write explicit expressions for the derivatives $\frac{\delta L}{\delta \partial_a \theta}$ in appendix \ref{d7stressvariation}.

We will now present all the components of the stress-energy tensor.

Many of the components are simple. For example, in the $S^3$ directions, and in the third Euclidean field theory direction, the only components are on the diagonal, and all are simply $-\int dz \, L = S_{D7}$.

The nontrivial components are in the $(z,t,x,y)$ subspace. Notice that, with one index up and one down, the energy-momentum tensor will not be symmetric, so we computed all sixteen components separately. We will also identify current components, $\langle J^t \rangle$, $\jx$ and $\jy$, whenever possible to simply the expressions.

We will now write all of the components of $\Theta^a_{~b}$ in the $(z,t,x,y)$ subspace. For notational simplicity, in what follows, we will not write the $\int dz$, which appears for every component. Primes will denote $\frac{\partial}{\partial z}$.

The components with upper index $t$ are
\beq
\begin{array}{ccccc}
\Theta^t_{~t} & = & -L - F_{xt} \frac{\delta L}{\delta F_{tx}} - F_{zt} \frac{\delta L}{\delta F_{tz}} & = & -L + E \frac{\partial L}{\partial E} + \langle J^t \rangle A_t' \bigskip \\
\Theta^t_{~x} & = & - F_{zx} \frac{\delta L}{\delta F_{tz}} - F_{yx} \frac{\delta L}{\delta F_{ty}} & = &  \langle J^t \rangle A_x' - \frac{B}{E} \left( \jx A_y' - \jy A_x' \right) \bigskip \\
\Theta^t_{~y} & = & -F_{xy} \frac{\delta L}{\delta F_{tx}} - F_{zy} \frac{\delta L}{\delta F_{tz}} & = &  B \frac{\partial L}{\partial E} + \langle J^t \rangle A_y' \bigskip \\
\Theta^t_{~z} & = & -F_{xz} \frac{\delta L}{\delta F_{tx}} - F_{yz} \frac{\delta L}{\delta F_{ty}} +\theta' \frac{\delta L}{\delta \partial_t \theta}& = &- A_x' \frac{\partial L}{\partial E} - A_y' \frac{\partial L}{\partial E_y} +\theta' \frac{\delta L}{\delta \partial_t \theta}\nonumber
\end{array}
\eeq

The components with upper index $x$ are
\beq
\begin{array}{ccccc}
\Theta^x_{~t} & = & -F_{zt} \frac{\delta L}{\delta F_{xz}} & = &  \jx A_t' \bigskip \\
\Theta^x_{~x} & = &- L - F_{tx} \frac{\delta L}{\delta F_{xt}} - F_{yx} \frac{\delta L}{\delta F_{xy}} - F_{zx} \frac{\delta L}{\delta F_{xz}} & = & - L + E \frac{\partial L}{\partial E} + B \frac{\partial L}{\partial B} + \jx A_x' \bigskip \\
\Theta^x_{~y} & = & - F_{zy} \frac{\delta L}{\delta F_{xz}} & = &  \jx A_y' \bigskip \\
\Theta^x_{~z} & = &- F_{tz} \frac{\delta L}{\delta F_{xt}} - F_{yz} \frac{\delta L}{\delta F_{xy}} +\theta' \frac{\delta L}{\delta \partial_x \theta} & = &  A_t' \frac{\partial L}{\partial E} + A_y' \frac{\partial L}{\partial B} +\theta' \frac{\delta L}{\delta \partial_x \theta} \nonumber
\end{array}
\eeq

The components with upper index $y$ are
\beq
\begin{array}{ccccc}
\Theta^y_{~t} & = & - F_{xt} \frac{\delta L}{\delta F_{yx}} - F_{zt} \frac{\delta L}{\delta F_{yz}} & = &  E \frac{\partial L}{\partial B} + \jy A_t' \bigskip \\
\Theta^y_{~x} & = & -F_{zx} \frac{\delta L}{\delta F_{yz}} - F_{tx} \frac{\delta L}{\delta F_{yt}} & = &  \jx A_y' \bigskip \\
\Theta^y_{~y} & = & -L - F_{xy} \frac{\delta L}{\delta F_{yx}} - F_{zy} \frac{\delta L}{\delta F_{yz}} & = &- L + B \frac{\partial L}{\partial B} + \jy A_y' \bigskip \\
\Theta^y_{~z} & = &- F_{tz} \frac{\delta L}{\delta F_{yt}} - F_{xz} \frac{\delta L}{\delta F_{yx}} +\theta' \frac{\delta L}{\delta \partial_y \theta} & = &  A_t' \frac{\partial L}{\partial E_y} - A_x' \frac{\partial L}{\partial B} +\theta' \frac{\delta L}{\delta \partial_y \theta} \nonumber
\end{array}
\eeq

The components with upper index $z$ are
\beq
\begin{array}{ccccc}
\Theta^z_{~t} & = & - F_{xt} \frac{\delta L}{\delta F_{zx}} & = & - \jx E \bigskip \\
\Theta^z_{~x} & = & - F_{tx} \frac{\delta L}{\delta F_{zt}} - F_{yx} \frac{\delta L}{\delta F_{zy}} & = &  \langle J^t \rangle E + \jy B \bigskip \\
\Theta^z_{~y} & = & -F_{xy} \frac{\delta L}{\delta F_{zx}} & = & - \jx B \bigskip \\
\Theta^z_{~z} & = & -L - \sum_{i=t,x,y} F_{iz} \frac{\delta L}{\delta F_{zi}} + \theta' \frac{\delta L}{\delta \theta'} & = & - L + \sum_{i=t,x,y} \langle J^i \rangle A_i' + \theta' \frac{\delta L}{\delta \theta'} \nonumber
\end{array}
\eeq
All quantities on the right-hand sides are evaluated on-shell. Those components with both indices in field theory directions ($t,\, x, \, y$) we can identify with the energy-momentum densities of the flavor fields, as explained above.

We would like to convert the components of $\Theta^a_{~b}$ to field theory quantities. In most cases, whether we can do so depends on whether we can perform the $z$ integration. Sometimes this is easy. For example, we know that $\int \, dz \, L = -S_{D7} = \Omega$, and $\int \, dz \, A_t'(z) = - \mu$, where $\mu$ is the $U(1)_B$ chemical potential. We thus have, for example, $\Theta^x_{~t} = \langle T^x_{~t} \rangle = - \mu \, \jx$. In some cases we can translate to SYM theory quantities without doing the $z$ integrals. For instance, terms with $\frac{\partial L}{\partial E}$ or $\frac{\partial L}{\partial B}$ multiplying $E$ or $B$ we interpret as contributions from the polarization tensor, as explained in the last subsection (and as will be verified explicitly, in certain limits, in section \ref{zeroTsection}). In the next subsection we will identify the $\Theta^z_{~i}$ components ($i=t,x,y$) with the rates of energy and momentum loss of the flavor fields. Notice also that the $\Theta^z_{~z}$ component is, up to the $\theta' \frac{\delta L}{\delta \theta'}$ term, identical to the Legendre transform in eq. (\ref{lteffaction}).

On the other hand, we have not found a field theory interpretation for the components $\Theta^t_{~z}$, $\Theta^x_{~z}$, and $\Theta^y_{~z}$, for which the $z$ integration is non-trivial. For many components ($\Theta^t_{~x}$, $\Theta^t_{~y}$, $\Theta^x_{~x}$, $\Theta^x_{~y}$, etc.), converting to SYM theory quantities requires integrating $A_x'$ and $A_y'$, and the field theory meaning is not immediately clear. We discuss the $z$ integration of $A_x'$ and $A_y'$ in appendix \ref{boundaryconditions}.

Finally, notice that $\Theta^t_{~x} = 0$ when $\jt=0$, partly because $\Theta^t_{~x}$ includes the expected polarization term $B M^{ty}$, and as explained in the last subsection, $M^{ty}=0$ when $\jt=0$. That $\Theta^t_{~x}$ vanishes when $\jt=0$ is easy to understand physically. As explained in appendix \ref{conductivity}, our system has two types of charge carriers, the charge carriers we introduced explicitly in the density $\jt$, but also charge carriers produced by pair production in the external electric field $E$. When $\jt=0$, we find a nonzero charge current $\jx$, coming from pair production. However, we expect the momentum current in the $x$ direction to vanish because charges produced in pairs will have zero \textit{net} momentum.

\section{Energy and Momentum Loss Rates}
\label{lossratessection}

As soon as we turn on any external fields in the plasma, the work done by the external forces will change the energy and momentum of our system at a constant rate. Without a mechanism for dissipation, a stationary solution is impossible, as the momentum (and energy) in the system will never stop increasing. The plasma is translation-invariant, so momentum can never really dissipate. However, in the probe limit $N_f \ll N_c$, the flavor degrees of freedom are very dilute relative to the much more abundant ${\cal N}=4$ SYM degrees of freedom. The current, momentum and energy densities of the flavor fields can be constant, but they will transfer energy and momentum density to the $\N=4$ plasma, which will thus gain energy and momentum at a constant rate. The conservation law for the full stress tensor is (see for example ref. \cite{Hartnoll:2007ih})
\beq
\label{rates}
\partial^{\mu} \langle T_{\mu \nu} \rangle = F_{\nu \rho} \langle J^{\rho} \rangle.
\eeq
In particular, for the spatially homogeneous solutions we considered, with only $F_{xt}=E$ and $F_{xy}=B$ nonzero, we have
\bea
\label{ratesexplicit}
\partial_t \langle T^t_{~t} \rangle & = & - E \langle J^x \rangle \\
\partial_t \langle T^t_{~x} \rangle & = & E \langle J^t \rangle + B \langle J^y \rangle \nonumber \\
\partial_t \langle T^t_{~y} \rangle & = & - B \langle J^x \rangle. \nonumber
\eea

From the point of view of the D7-branes' stress-energy tensor, these loss rates are reflected in the appearance of an IR divergence, very similar to what happened for the dragging string solution of refs. \cite{Herzog:2006gh,Gubser:2006bz}. To be more specific, $\Theta^t_{~t}$, $\Theta^t_{~x}$ and $\Theta^t_{~y}$ have divergences from the $z=z_H$ endpoint of integration (in addition to any expected UV divergences from the $z=0$ endpoint, which can be cancelled with counterterms).

If we regulate these IR divergences by only including those parts of spacetime that had time to communicate with the boundary within a time $t$, we find that the divergences are linear in $t$ and the coefficients can be interpreted as loss rates. Let us demonstrate this quantitatively. We simply need to expand the expressions from the previous section and study the behavior close to the horizon. Let $z = z_h - \epsilon$. We find that in all three cases the integrand has terms that diverge as $\frac{1}{\epsilon}$. The loss rate should be proportional to the coefficient of that pole, but we must fix the exact proportionality constant. To do so, we can compare the $\frac{1}{\epsilon}$ poles to the time required for a ray of light to propagate from $z=z_h - \epsilon$ out to the boundary, which is given by
\begin{equation}
t_{light} = \int_0^{z_H -\epsilon} dz \, \sqrt{\frac{g_{zz}}{|g_{tt}|}}.
\end{equation}
The integrand in $t_{light}$ itself diverges as $\frac{1}{\epsilon}$, producing the $\frac{1}{\epsilon}$ pole. We can thus extract finite rates by identifying the divergent parts of $\Theta^t_{~i}$ as $t_{light}$ times the rate. The rates so obtained are in perfect agreement\footnote{The limit of zero temperature but finite density, with $E>B$, is more subtle. In this case, the system is not really stationary, as we review in more detail in appendix \ref{zeroTformulas}. Indeed, in that case the loss rate seems to acquire more terms. Of course what really happens at zero temperature is that the real physical loss rate is no longer time-independent, as the charge carriers continuously accelerate. The loss rates from $\Theta^{z}_{~i}$ on the other hand still agree with the hydrodynamic expectation.} with eq. (\ref{rates}).

The right-hand sides in eq. (\ref{ratesexplicit}) are closely related to $\Theta^{z}_{~i}$ with $i = t,x,y$. Specifically, the $\Theta^z_{~i}$ differ only by a factor of the $z$ integration, $\int_0^{z_H} dz = z_H =  \frac{\sqrt{2}}{\pi T}$, from the loss rates on the right-hand sides in eq. (\ref{ratesexplicit}).

In ref. \cite{Gubser:2008vz}, the response of the SYM theory plasma to a moving source, such as a heavy quark, was computed using the usual AdS/CFT framework, that is, by introducing a probe source in $AdS_5$, such as a long string, computing the back-reaction on the metric to linear order in the perturbation caused by the probe, and then extracting the SYM theory stress-energy tensor from the asymptotic form of the back-reacted metric. One of the main results of ref. \cite{Gubser:2008vz} was that the non-conservation of the stress-energy tensor (the right-hand sides of eq. (\ref{ratesexplicit})) was determined by the ``$zi$'' components of the probe's stress-energy tensor. We have reached the same conclusion using the Hamiltonian AdS/CFT framework, which confirms the equivalence of the two approaches.

Given the above loss rates, we can construct two IR-safe quantities,
\begin{eqnarray}
\nonumber
I_1 &=& E \langle T^t_{~y} \rangle - B \langle T^t_{~t} \rangle \\
I_2 &=& \<J^x \> \langle T^t_{~x} \rangle + \< J^y \> \langle T^t_{~y} \rangle +\< J^t \> \langle T^t_{~t} \rangle.
\end{eqnarray}
Both $I_1$ and $I_2$ are free of IR divergences, so the quantities they represent on the field theory side are constant as long as the currents are time-independent. We can write these two IR-safe objects as the $t$ components of the currents
\begin{eqnarray}
I_1^{\mu} & =& \epsilon^{x_3 \nu \rho \sigma} \langle T^{\mu}_{~\nu} \rangle F_{ \rho \sigma} \\
I_2^{\mu} & = & \langle T^{\mu}_{~\nu} \rangle \langle J^{\nu} \rangle
\end{eqnarray}
where $x_3$ refers to the third Euclidean spatial direction in the field theory. We can easily show that these two currents are conserved. The divergence of $I_1^{\mu}$ is proportional to $F \wedge F$ and so vanishes in topologically trivial background fields, such as our orthogonal $E$ and $B$. (An interesting generalization of our results, and those of ref. \cite{Karch:2007pd,O'Bannon:2007in}, would be to the case with non-trivial $\vec{E} \cdot \vec{B}$.) Using $\partial_{\mu} \langle T^{\mu \nu} \rangle = F^{\nu \rho} \< J_{\rho} \>$, we find that the divergence of  $I_2^{\mu}$ is $F_{\mu \nu} \< J^{\mu} \rangle \langle J^{\nu} \>$, which vanishes identically.

As conservation of $I_1^{\mu}$ seems to be a peculiarity of the background we chose, we will focus on the interpretation of $I_2^{\mu}$, which is conserved whenever the currents are stationary. Indeed, $I_2^{\mu}$ has a simple interpretation: it represents the mass-energy current density 4-vector of the fluid as measured by inertial observers with a 4-velocity proportional to $\< J^{\nu} \>$, (see e.g. section 4.2 of ref. \cite{Wald:1984rg}). At least in the limit where the current is only carried by quasi-particles, we can understand this result in simple terms. In the rest frame of the charge carriers, the force due to the external fields does not lead to any work, as $W= \int \vec{F} \cdot \vec{ds} = 0$ (the charge carriers are at rest and the fluid moves). Such an effect has already been seen in comparing the results for the dragging string of refs. \cite{Herzog:2006gh,Gubser:2006bz} to the calculation of ref. \cite{Liu:2006he}, both describing a quark pulled by an electric field and moving at constant speed through the plasma. In the rest frame of the fluid, a quark that has been moving for a long time builds up an energy proportional to the distance traveled \cite{Herzog:2006gh,Gubser:2006bz}. In the rest frame of the quark, the same quark has a time-independent, finite energy.

\section{The Limit of Zero Temperature}
\label{zeroTsection}

To extract transport coefficients from currents we need a stationary system. For generic external fields, our system is not stationary, as visible from the loss rates discussed above. In certain limits, however, the loss rates may vanish or be higher order in the external fields. In this section, we consider such limits, and check that our results agree with the expectations of the SYM theory.

A special case in which all the loss rates vanish is when all the currents vanish, so that we are analyzing equilbrium physics. As we discussed in section \ref{polarizationsection}, polarization effects due to the external fields may still produce non-vanishing momentum currents even when the currents $\jx$ and $\jy$ are zero. At zero temperature, the two cases where we can set all currents to zero consistently are $E=0$ and $B>E$. From the supergravity perspective, these are the only cases where $\jx = \jy = 0$ and the D7-brane action remains real for all $z$.

A third scenario in which the loss rates can be neglected, even for $E>B$, is when $E\sim \epsilon$ and $B \sim \epsilon$ for some very small $\epsilon$. As $\epsilon$ carries dimension what this scaling really means is that we can neglect the loss rates on time scales smaller than $1/\sqrt{\epsilon}$. In this case, the loss rates at zero temperature are of order $\epsilon^{5/2}$, whereas the leading contribution to the momentum density is of order $\epsilon^2$, hence to leading order the loss rates are negligible. This is due to the somewhat peculiar scaling $\langle J_x \rangle \sim E^{3/2}$ of the current at zero temperature (see appendix \ref{zeroTformulas}). At finite temperature, where the current has typical Ohmic form (linear in the electric field), we have not been able to identify an interesting scenario where the loss rate can be neglected. All the examples we will present thus involve the specialization to zero temperature.

\subsection{$E=0$, $B$ and $\jt$ nonzero}

As a first example where the loss rates vanish, we simply set $E=0$, so that $\jx = \jy = 0$, and the system is in equilibrium. In our D7-brane solution, only $A_t(z)$ and $B$ are nonzero. From our result for $\Theta^t_{~t} = \langle T^t_{~t} \rangle$, we find, when $E=0$,
\beq
\Theta^t_{~t} = - \int dz \, \left( L - \langle J^t \rangle A_t' \right )
\eeq
which, using $\int dz \, L = - S_{D7} = \Omega$, immediately implies
\beq
\label{zeroEzeroTthermo}
\O = -\langle T^t_{~t} \rangle - \mu \langle J^t \rangle.
\eeq
At zero temperature we do not expect a horizon contribution to $\langle T^t_{~t} \rangle$, hence $\langle T^t_{~t} \rangle$ should accurately represent the energy density. Eq. (\ref{zeroEzeroTthermo}) is indeed the correct thermodynamic relation between the energy density $-\langle T^t_{~t} \rangle$ and the grand canonical free energy density.

Notice that our general result for $\Theta^t_{~t}$ remains the same at finite $T$. At finite temperature eq. (\ref{zeroEzeroTthermo}) is not the right relation between free energy and energy density, however. We are missing a $T s$ contribution, where $s$ in the entropy density, for which we would have to calculate the change in horizon area due to the back-reaction. This is in perfect analogy to what we found in the toy model of section \ref{toymodel}: the stress-energy tensor of the probe completely misses the entropy contribution.

\subsection{$E=0$, $B=0$, $\jt$ and $\jx$ nonzero}

At zero temperature, if all external fields vanish then the system has no dissipation. We may of course introduce $\langle J_t \rangle$, but also constant $\langle J_x \rangle$ and $\langle J_y \rangle$. At zero temperature the system is boost-invariant, so we can always boost to a rest frame in which only $\langle J_t \rangle$ is non-zero. Without the electric field, nothing distinguishes $x$ from $y$, so for simplicity we only present the result for $\langle T^t_{~x} \rangle$, since the expression for $\langle T^t_{~y} \rangle$ is identical.

At zero temperature we have $g_{xx}=|g_{tt}|$, while $g_{zz} = \frac{1}{z^2} + (\theta')^2$ is generically nontrivial. Fortunately, as we show in appendix \ref{d7stressvariation} (see eq. (\ref{ttxfromatp})), $\langle T^t_{~x} \rangle$ does not depend on $g_{zz}$:
\beq
\langle T^t_{~x} \rangle = - \langle J_x \rangle \int dz \, A_t'
\frac{g_{xx}}{|g_{tt}|} = - \langle J_x \rangle \int dz A_t' = \langle J_x \rangle \mu.
\eeq
With this we can determine the heat current,
\beq
\langle Q_x \rangle = \langle T^t_{~x} \rangle- \mu \langle J_x \rangle = 0.
\eeq

This result is actually dictated by Lorentz invariance, as can be easily seen via field theory arguments.
In the rest frame of the charges, with density $\langle J^t \rangle^{rest} = \rho$, energy density $-\langle T^t_{~t}\rangle = \epsilon^{rest}$, and pressure $\langle T^x_{~x}\rangle=\langle T^y_{~y} \rangle=P^{rest}$,
we know that $\langle J_x \rangle^{rest} = \langle T^t_{~x} \rangle =0$.
Boosting with velocity $v$ and boost parameter $\gamma^{-1}=\sqrt{1-v^2}$, we find
\beq
\langle J_x \rangle=\langle J^x \rangle = v \gamma \rho, \qquad \langle J^t \rangle = \gamma \rho,
\eeq
and hence the momentum density in the \textit{boosted} frame is
\beq
\langle T^t_{~x} \rangle = (\epsilon^{rest} + P^{rest}) v \gamma^2 = \langle J_x \rangle \mu,
\eeq
where in the last step we used the equilibrium relation $\epsilon^{rest} + P^{rest} = \mu^{rest} \rho$ and
we needed to absorb one power of $\gamma$ into $\mu= \mu^{rest} \gamma$. In a relativistic theory $\mu$ is best viewed as the constant expectation value for the $t$ component of a background gauge field $A_{\mu}$. As such it transforms nontrivially under Lorentz boosts. That our supergravity result correctly reproduces the SYM theory expectation is encouraging.

\subsection{$B > E$, $\langle J^t \rangle = 0$}
\label{BGE}

If the density of charge carriers is zero and the magnetic field is larger than the electric field, then $\jx = \jy = 0$ and the loss rates vanish (see the discussion in appendix \ref{zeroTformulas}). We do not expect to find any transport coefficients at zero temperature and density, but we should be able to understand the stress-energy tensor in terms of polarizations. In particular, we expect $\langle T^t_{~y} \rangle = \Theta^t_{~y} = M^{tx} B$. (Recall that with $\jt=0$ we have $\Theta^{t}_{~x} = - M^{ty} B = 0$, as explained at the end of section \ref{d7stressenergytensor}.)

To verify that $\langle T^t_{~y} \rangle = M^{tx} B$, we will focus on the meta-stable $\theta=0$ solution representing massless flavor fields. From our result for the stress-energy tensor with $\jt=0$ we have
$\Theta^t_{~y} =  B \int dz \frac{\partial L}{\partial E}$. When $\jx=\jy=0$ and $\theta=0$, the boundary terms in eq. (\ref{bulkpolarizationequation}) vanish, so we can identify $M^{tx} = \int dz \, \frac{\partial L}{\partial E}$ and hence $\Theta^t_{~y} = M^{tx}B$. As expected, in this limit only the polarization term contributes to the momentum density.

\subsection{$ E \gg B$ but both $E$ and $B$ small, $\langle J^t \rangle = 0$}
\label{EGGB}

Finally, if we assume $E \gg B$, but that both $E$ and $B$ are of the order of some small parameter $\epsilon$, and assume again that $\langle J^t \rangle = 0$, then at zero temperature all loss rates will be of order $\epsilon^{5/2}$.
If we work to order $\epsilon^2$, all quantities will be free of IR divergences and we should again find only polarization contributions to the stress-energy tensor. We will also set $\theta(z) = 0$ again, for simplicity.

With $\jt = 0$ we have $\jy=0$ and again $\Theta^t_{~x} = 0$. We would like to confirm that $\Theta^t_{~y} = M^{tx} B$ to order $\epsilon^2$. Now in eq. (\ref{bulkpolarizationequation}) the $A_x'$ boundary term is non-vanishing, so
\beq
M^{tx} =  \int dz \, \frac{\partial L}{\partial E} + \left [ \jx \frac{\partial A_x}{\partial E} \right ]^{z_H}_0.
\eeq
$\frac{\partial L}{\partial E}$ will be proportional to $E$ (the $\frac{\partial}{\partial E}$ acts on the $E^2$ term in eq. (\ref{a2def})). We also have $\left . \frac{\partial A_x}{\partial E}\right |_0^{z_H} = \frac{\partial}{\partial E} \int_0^{z_H} dz A_x'$, where the leading $E$ dependence of $A_x'$ is contained in $\jx$ (see eq. (\ref{ataxsol})). As explained in appendix \ref{zeroTformulas}, at zero temperature $\jx \sim E^{3/2}$ so in our limit $\jx \sim \epsilon^{3/2}$. The boundary term is thus of order $\jx \frac{\partial \jx}{\partial E} \sim \epsilon^2$, so $B \sim \epsilon$ times the boundary term is order $\epsilon^{3}$, and thus we find
\beq
\langle T^t_{~y} \rangle = \Theta^t_{~y} =  B \int dz \, \frac{\partial L}{\partial E} = M^{tx} B + O\left( \epsilon^{3} \right).
\eeq
Up to order $\epsilon^2$, only the polarization term contributes to the momentum density.

\section{Conclusion}
\label{conclusion}

We began with the intention of extracting thermoelectric transport coefficients from the heat currents of $\mathcal{N} = 4$ SYM theory with flavor fields. However, we have found that the probe brane approximation does not adequately separate the physics of the flavor charge carriers from the physics of the $\N=4$ SYM plasma.  As a result, the loss rates of the charge carriers provide infinite contributions to the heat currents: for this supergravity background, and for our particular ansatz for D7-brane worldvolume fields, stationary solutions that would allow us to define the time-independent thermoelectric coefficients do not exist.

Using the natural time-scale of the problem $t_{light}$, we were able to extract the loss rates and compare them to field theory expectations, to which they conformed perfectly.  Also, we were able to calculate the heat currents in specific zero-temperature cases and confirm that they had the expected structure in terms of currents and polarizations.

Another important lesson that we learned along the way is that the probe brane stress-energy tensor alone does not give the full order $N_f/N_c$ contribution to the energy of the system. In order to calculate the energy to this order, the back-reaction of the probes on the horizon must be included.

A possible next step would be to include flavor fields in the fluid dynamical formulation of ref. \cite{Bhattacharyya:2008jc}. By looking at the back-reaction in the form of a slowly-varying horizon, the SYM theory thermoelectric transport coefficients could be extracted.

\section*{Acknowledgements}
We thank Dam Son for useful discussions. We also
thank Rob Myers for illuminating email exchanges and for explaining to us
the results of ref. \cite{myersun}. AK would like to thank the Max Planck Institute for Physics in Munich and in particular Dieter L\"ust for hospitality while parts of this work were completed. This work was supported in part by the U.S. Department of Energy under Grant No. DE-FG02-96ER40956. The work of ET was also supported in part by DE-FG02-00ER41132. The work of A.O'B. was also supported in part by the Cluster of Excellence ``Origin and Structure of the Universe.''

\begin{appendix}

\section{The Conductivity}
\label{conductivity}

In this appendix we will review the results of refs. \cite{Karch:2007pd,O'Bannon:2007in} for the conductivity tensor of the flavor fields, as computed using probe D7-branes in AdS/CFT. We then briefly discuss the boundary conditions on the D7-brane worldvolume gauge fields, and some subtleties about the zero-temperature limit of the result for the conductivity.

In the SYM theory, we expect $\jx$ and $\jy$ to be fixed once we choose $E$, $B$ and (in the canonical ensemble) $\langle J^t \rangle$. The main observation of refs. \cite{Karch:2007pd,O'Bannon:2007in} was that only one choice of $\jx$ and $\jy$ allows for the gauge fields and the on-shell action to be real for all values of $z$ between the horizon and the boundary. More specifically, the three functions $\xi$, $\chi$ and $a$ defined above in eq. (\ref{xichiadef}) all have a zero between the horizon and the boundary, and the only way for the action and the gauge fields to remain real is if all three functions pass through zero simultaneously. We thus get three equations ``for free.'' The location of the zero, which we denote $z_*$, is given by $\xi = 0$. In terms of the dimensionless quantities
\beq
e = \frac{1}{2} (2\p\a') E z_H^2 = \frac{E}{\frac{\p}{2}\sqrt{\lam}T^2}, \qquad b = \frac{1}{2}(2\p\a')B z_H^2 = \frac{B}{\frac{\p}{2} \sqrt{\lam}T^2}
\eeq
the explicit form of $z_*$ is
\bea
\label{zstardef}
\frac{z_*^4}{z_H^4} & = & e^2 - b^2 + \sqrt{(e^2 - b^2)^2 + 2 (e^2 + b^2) + 1}\\ & & - \sqrt{ \left ( (e^2 - b^2) + \sqrt{(e^2 - b^2)^2 + 2 (e^2 + b^2) + 1} \right )^2 - 1}.  \nonumber
\eea
At that point, $g_{xx}^2(z_*) = \p^4 T^4 {\cal F}(e,b)$ where
\beq
\label{calF}
{\cal F}(e,b) = \frac{1}{2} \left ( 1+ e^2 - b^2 + \sqrt{(e^2 - b^2)^2 + 2 (e^2 + b^2) + 1}\right ).
\eeq
The remaining two equations, $a = \chi = 0$, then fix the values of the currents,
\begin{subequations}
\label{cxcy}
\beq
\label{cx}
\< J^x \> = \frac{E g_{xx}}{g_{xx}^2 + (2\p\a')^2 B^2} \sqrt{(g_{xx}^2 + (2\p\a')^2 B^2) \N^2 (2\p\a')^4 g_{xx} \cos^6 \th(z_*) + (2\p\a')^2 \<J^t\>^2}
\eeq
\beq
\label{cy}
\< J^y \> = -\frac{(2\p\a')^2 \<J^t\>B}{g_{xx}^2 + (2\p\a')^2 B^2} E
\eeq
\end{subequations}
\noi with all functions of $z$ evaluated at $z_*$. Converting to field theory quantities, we have the conductivity,

\begin{subequations}
\label{d7sigma}
\beq
\s_{xx} = \sqrt{\frac{N_f^2 N_c^2 T^2}{16 \p^2} \frac{{\cal F}^{3/2}}{b^2 + {\cal F}}\cos^6 \th(z_*) + \frac{\rho^2 {\cal F}}{(b^2 + {\cal F})^2}},
\eeq
\beq
\s_{xy} = \frac{\rho b}{b^2 + {\cal F}},
\eeq
\end{subequations}
where $\rho$ is defined similarly to $e$ and $b$,
\beq
\rho = \frac{\< J^t \>}{\frac{\p}{2} \sqrt{\lam}T^2}.
\eeq

Notice that $\sigma_{xx}$ clearly has two terms adding in quadrature, and that even if we set $\jt=0$ (so $\rho=0$), we find a nonzero $\sigma_{xx}$. Even without the density of charge carriers $\jt$, if we impose an electric field we will see a current $\jx$, from the $\cos \theta(z_*)$ term in $\sigma_{xx}$. In other words, the system has some other charge carriers besides those in $\jt$. What is the source of these other charge carriers? The claim of ref. \cite{Karch:2007pd} was that they come from pair production in the external electric field. The primary piece of evidence is the behavior of $\theta(z_*)$. As explained at the end of section \ref{d7solution}, $\theta(z_*)=0$ maps to $m=0$ in the SYM theory, and $\theta(z_*) \rightarrow \frac{\pi}{2}$ maps to $m \rightarrow \infty$. In the former limit we expect pair production to be maximal, and indeed the current coming from pair production is a maximum: $\cos \theta(z_*) = 1$. In the other limit we expect pair production to be suppressed, and indeed we see $\cos \theta(z_*) \rightarrow 0$. Notice also that the pair-produced charges do not contribute to the Hall current: $\sigma_{xy}$ is proportional to $\jt$.

In appendix \ref{zeroTformulas} we discuss the zero-temperature limit of the conductivity.

\subsection{Boundary Conditions on Gauge Fields}
\label{boundaryconditions}

The D7-brane worldvolume gauge fields have the following asymptotic behavior:
\begin{subequations}
\label{asymptotic}
\beq
A_t(z) = \m - \frac{1}{2} \frac{\< J^t \>}{\N (2\p \a')^2} z^2 + O(z^4),
\eeq
\beq
A_x(z) = -E t + c_x + \frac{1}{2} \frac{\< J^x \>}{\N (2 \p \a')^2} z^2 + O(z^4),
\eeq
\beq
A_y(z) = B x + c_y + \frac{1}{2} \frac{\< J^y \>}{\N (2 \p \a')^2} z^2 + O(z^4),
\eeq
\end{subequations}
where $c_x$ and $c_y$ are constants.

In the usual AdS/CFT recipe, we fix the leading coefficients and then ``integrate into the bulk,'' where some boundary condition then fix the solutions completely, and hence fix the values of the sub-leading coefficients. This process maps onto the field theory process in which, once we choose the parameters in the Lagrangian, the dynamics of the theory fixes expectation values of operators. From this point of view, our calculation of the conductivity looks strange. We solved for the field strengths (see eqs. (\ref{ataxsol}) and (\ref{aysol})) and then imposed reality of the D7-brane action to fix the values of $\jx$ and $\jy$. Implicitly, we must be imposing some boundary condition on the gauge fields somewhere in the bulk of $AdS_5$. In this section we briefly clarify the boundary conditions on the gauge fields.

We first consider $A_t$. The leading, non-normalizable constant we identify as the $U(1)_B$ chemical potential, $\mu$. In this case, the background geometry forces a boundary condition on $A_t(z)$: the Killing vector corresponding to time translations becomes degenerate at the horizon, hence for the gauge field to remain well-defined as a one-form we must take $A_t(z_H)=0$ \cite{Kobayashi:2006sb}. Notice that we are then also restricted to the subset of gauge transformations that leaves this boundary condition invariant, which is why $\mu$ has physical meaning (we cannot simply gauge it away).

$A_x$ and $A_y$ are more subtle. For these, the leading, non-normalizable terms include $-Et$ and $Bx$, which give rise to the gauge-invariant field strengths $F^{tx} = E$ and $F^{xy} = B$, but the leading terms also include the constants $c_x$ and $c_y$. We will choose a gauge in which these are zero. What then is the IR boundary condition that fixes $\jx$ and $\jy$? The field strengths $A_t'(z)$ and $A_y'(z)$ are of course gauge-invariant, as are their integrals,
\beq
\int_0^{z_H} dz \, A_x'(z) = h_x(z_H), \qquad \int_0^{z_H} dz \, A_y'(z) = h_y(z_H).
\eeq
Here the background geometry does not force any particular values for $h_x(z_H)$ and $h_y(z_H)$ upon us. Instead, we are implicitly choosing these constants to produce precisely the values of $\jx$ and $\jy$ that keep the action real for all $z$. The values of $h_x(z_H)$ and $h_y(z_H)$ corresponding to these values of $\jx$ and $\jy$ could be computed explicitly by performing the above integrals, using the solutions in eqs. (\ref{ataxsol}) and (\ref{aysol}) and the values of $\jx$ and $\jy$ in eq. (\ref{cxcy}). Crucially, however, these integrals diverge in our setup.

Many components of the stress-energy tensor that we computed in section \ref{setensorsection} can only be translated to field theory quantities using $\int dz A_x'$ and $\int dz A_y'$. As we discussed in section \ref{lossratessection}, the divergences in these integrals are related to loss rates in the SYM theory, and indeed we expect some components of the stress-energy tensor to be sensitive to the loss rates. The conductivity does not depend on the values of $\int dz A_x'$ and $\int dz A_y'$, however: it is insensitive to the loss rates.

\subsection{The Conductivity at Zero Temperature}
\label{zeroTformulas}

Here we study the zero-temperature limit of the conductivity, eq. (\ref{d7sigma}), in detail.

The zero-temperature limit of the quantity ${\cal F}$ defined in eq. (\ref{calF}),
\beq
{\cal F}(e,b) = \frac{1}{2} \left ( 1+ e^2 - b^2 + \sqrt{(e^2 - b^2)^2 + 2 (e^2 + b^2) + 1}\right )
\eeq
is subtle. Recall that an inverse power of $T^2$ is hidden in the definitions of $e$ and $b$, so these quantities grow large as $T$ becomes small. Correspondingly, the term under the square root is dominated by $\sqrt{(e^2-b^2)^2} = | e^2 - b^2|$. Due to the absolute value, we must take care to distinguish between the $B>E$ and the $B<E$ cases.

For $B>E$, the leading order $1/T^4$ piece in ${\cal F}$ vanishes, and ${\cal F}$ goes to a constant: ${\cal F}=
B^2/(B^2-E^2)$. In this limit, $\sigma_{xx}$ vanishes and $\sigma_{xy} = \< J_t \>/B$ as required by Lorentz invariance, since for $B>E$ we can boost back to a system with $E=0$, hence no spatial currents and $B$-field only.

On the other hand, for $E>B$ we find ${\cal F} = e^2 -b^2$. With this both $\sigma_{xx}$ and $\sigma_{xy}$ have a finite limit. Specializing for example to the zero-mass, zero-density case of $\theta=\rho=0$, we find $\< J_x \> \sim E^{3/2}$. The system has no linear (Ohmic) current, having instead a current that scales as a fractional power of the electric field. Indeed, this power of $E$ is dictated by the scale invariance of the zero-density, zero-mass theory. The strong electric field polarizes the medium so that even in the absence of charge carriers the medium can conduct at zero temperature.

One note of caution: these finite zero-temperature conductivities can be misleading. For example, in the limit of large mass and large density, where $\sigma_{xx}$ is completely dominated by the second term under the square root in eq. (\ref{d7sigma}), at zero $T$ we find $\sigma_{xx}=\< J_t \> / E$, or in other words $\< J_x \> = \< J_t \>$. In this
case the dynamics is easy to understand: the finite density of charge carriers accelerates forever, and hence comes
closer and closer to the speed of light. Both $\< J_x \> \sim v \gamma$ and $\< J_t \> \sim \gamma$ diverge individually
as the boost factor $\gamma$ increases. Their ratio, however, approaches a constant. In this case the finite conductivity does not represent a truly stationary system. The case with zero $\< J_t \>$, however, seems to be truly stationary.

\section{Stress-Energy Tensor from Variation of the Action}
\label{d7stressvariation}

Our goal in this appendix is to compute the stress-energy tensor of the D7-branes directly, by varying the D7-branes' action with respect to the background metric. We do so for two reasons: first, to check our calculation via the Noether-derived form, eq. (\ref{noetherstressenergy}), and second, to provide explicit formulas that may be useful for future calculations of the back-reaction of the D7-branes.

Let us first fix our notation. In the rest of the paper, $g_{ab}$ denoted the induced metric of the D7-brane. In this appendix only, $g_{ab}$ will denote the metric of the background spacetime in which the D-brane is embedded. For the particular ansatz in section \ref{d7solution}, in which the only nontrivial worldvolume scalar was $\theta(z)$, the distinction was irrelevant except for $g_{zz}$. The induced metric had $g_{zz} = \frac{1}{z^2} + \theta'^2$, but was otherwise identical to the background metric. The distinction between the two is crucial for computing the stress-energy tensor. Physically, we want to compute the D7-brane's response to a variation of the background metric, not to a variation of its induced metric. Also, in this appendix only, we will denote multiplication by $2\pi \alpha'$ with a tilde.  For example, $\tilde{F}_{ab} = (2 \pi \alpha') F_{ab}$.

We begin with an expression for the D7-branes' energy-momentum density,
\beq
\label{gravU}
U^{ab} = \frac{2}{\sqrt{-g_{D7}}} \frac{1}{2} \( \frac{\delta S_{D7}}{\delta g_{ab}} + \frac{\delta S_{D7}}{ \delta g_{ba}} \),
\eeq
where we have taken a symmetrized derivative with respect to the external metric $g_{ab}$ to guarantee a symmetric tensor. Lowering one of the indices in eq. (\ref{gravU}), and plugging into eq. (\ref{d7setensor}), we find
\beq
\label{Tvar}
\Theta^a_{~c} =  \int dz d^3\alpha ~ \( \frac{\delta S_{D7}}{\delta g_{ab}} + \frac{\delta S_{D7}}{ \delta g_{ba}} \) g_{bc}.
\eeq
Let us introduce some more notation. Let
\beq
m_{ab} \equiv \, g_{cd} \, \frac{\partial X^{c}}{\partial \zeta^a} \frac{\partial X^{d}}{\partial \zeta^b} + \tilde{F}_{ab}, \qquad m \equiv \textrm{det} (m_{ab}),
\eeq
where the first term in $m_{ab}$ is the induced metric of the D-brane. The $X^{a}$ represent the worldvolume scalars and the $\zeta^a$ represent the worldvolume coordinates. We then have simply
\beq
S_{D7} = - N_f T_{D7} \int d^8 \zeta~ \sqrt{ -m } \, .
\eeq
We can then use the chain rule
\beq
\frac{\delta m}{\delta g_{ab}} = \frac{\delta m}{\delta m_{cd}}\frac{\delta m_{cd}}{\delta g_{ab}}
\eeq
and the identities
\beq
\frac{\delta m_{cd}}{\delta g_{ab}} =  \frac{\partial X^{a}}{\partial \zeta^c} \frac{\partial X^{b}}{\partial \zeta^d}, \qquad \frac{\delta m}{\delta m_{cd}} = m \mi_{dc}
\eeq
to compute the variation with respect to the external metric:
\beq
\frac{\delta S_{D7}}{\delta g_{ab}} = \frac{N_f T_{D7} }{2} \frac{1}{\sqrt{-m}} \frac{\delta m}{\delta g_{ab}}
=\frac{N_f T_{D7} }{2}\frac{ m \mi_{dc}}{\sqrt{-m}} \, \frac{\partial X^{a}}{\partial \zeta^c} \frac{\partial X^{b}}{\partial \zeta^d} .
\eeq
Our expression for the variation of the D7-brane action, eq. (\ref{Tvar}), becomes
\beq
\Theta^{a}_{~c} = \int dz d^3 \alpha ~ \frac{N_f T_{D7}}{2} \frac{m \mi_{de}}{\sqrt{-m}} \( \frac{\partial X^{a}}{\partial \zeta^e} \frac{\partial X^{b}}{\partial \zeta^d} + \frac{\partial X^{b}}{\partial \zeta^e} \frac{\partial X^{a}}{\partial \zeta^d}\) g_{bc}
\eeq
which is the main result of this appendix, and does not depend on our particular ansatz for the worldvolume fields. Indeed, formally, upon changing the dimension and the tension, this formula is valid for any D-branes described by the Born-Infeld action in a background for which the dilaton and axion are trivial. A nontrivial NS two-form would simply add a term to our definition of $m_{ab}$.

To compute $\Theta^a_{~b}$ for our particular worldvolume fields, we need to write the matrix $m_{ab}$ explicitly and compute the elements of its inverse. To write $m_{ab}$ explicitly, we will need an explicit metric for the $S^3$,
\beq
ds^2_{S^3} = d\alpha_1^2 + \sin^2 \alpha_1 \, d\alpha_2^2 + \sin^2 \alpha_1\sin^2 \alpha_2 \, d\alpha_3^2.
\eeq
For our ansatz, $m_{ab}$ is
\beq
\( \begin{array}{cccccccc} g_{tt} & -\E & 0 & -\At & 0 & 0 & 0 & 0 \\ \E & \gx & \B & -\Ax & 0 & 0 & 0 & 0 \\
0 & -\B & \gx & -\Ay & 0 & 0 & 0 & 0 \\ \At & \Ax & \Ay & (g_{zz}+\theta'^2) & 0 & 0 & 0 & 0 \\ 0 & 0 & 0 & 0 & \gx & 0 & 0 & 0 \\ 0 & 0 & 0 & 0 & 0 & \cos^2 \theta & 0 & 0 \\ 0 & 0 & 0 & 0 & 0 & 0 & \cos^2 \theta \sin^2 \alpha_1 & 0 \\ 0
& 0 & 0 & 0 & 0 & 0 & 0 & \cos^2 \theta \sin^2 \alpha_1 \sin^2 \alpha_2
\end{array} \)
\eeq
which results in
\beq
\sqrt{-m} = \sqrt{\gx} \cos^3 \theta \sin^2 \alpha_1 \sin \alpha_2 \sqrt{d(z)},
\eeq
where we have defined the shorthand notation
\beq
\begin{array}{ccc}
d(z) & = & - g - (2 \p \a')^2 \frac{1}{2} \, g \, F^2 - (2 \p \a')^4 \frac{1}{4} \left(  F \wedge F \right)^2 \bigskip \\ & = & \gt \gx^2 (g_{zz}+\theta'^2) + (g_{zz}+\theta'^2) (\B^2 \gt - \E^2 \gx) \smallskip \\ & &  - \gx^2 (\At)^2 + \, \gt \gx \( (\Ax)^2 + (\Ay)^2 \) - \( \Ay \E - \At \B \)^2 \end{array}
\eeq
where in the first line $g$ is the determinant of the induced D7-brane metric in the $(z,t,x,y)$ subspace (\textit{i.e.} minus the first term in the second line).

Let us explicitly compute one example, $\Theta^t_{~x}$, for which we have
\beq
\Theta^{t}_{~x} = \int dz d^3 \alpha ~ \frac{N_f T_{D7}}{2} \frac{1}{\sqrt{-m}} \( m \mi_{xt} + m \mi_{tx} \) g_{xx},
\eeq
hence we compute
\begin{subequations}
\beq
m \mi_{xt} = g_{xx} \cos^6 \theta(z) \sin^4 \alpha_1 \sin^2 \alpha_2 \left( - g_{xx} \At \Ax - \E g_{xx} (g_{zz}+\theta'^2) - \E (\Ay)^2 + \B \At \Ay \right) \nonumber
\eeq
\beq
m \mi_{tx} = g_{xx} \cos^6 \theta(z) \sin^4 \alpha_1 \sin^2 \alpha_2 \left( - g_{xx} \At \Ax + \E g_{xx} (g_{zz}+\theta'^2) + \E (\Ay)^2 - \B \At \Ay \right). \nonumber
\eeq
\end{subequations}
 With these expressions, and performing the integration over the $S^3$, we find
\beq
\Theta^t_{~x} =  -\int dz~ \mathcal{N} \cos^3 \theta \, \gx^{5/2} \frac{\At \Ax}{\sqrt{d(z)}}
\eeq
which, using the definition of $\jx$ in eq. (\ref{currentdefinition}), can also be written as
\beq
\label{ttxfromatp}
\Theta^t_{~x} = -\langle J_x \rangle \int dz~ A_t' \, \frac{\gx}{\gt}.
\eeq
The same procedure then gives, for example,
\begin{eqnarray}
\nonumber
\Theta^{t}_{~y} &=& - \int dz~ \mathcal{N} \cos^3 \theta \gx^{3/2} \frac{ (g_{zz}+\theta'^2) \B \E + \gx \At \Ay}{\sqrt{d(z)}}\\
\Theta^t_{~t}&=& - \int dz~ \mathcal{N} \cos^3 \theta g^{1/2}_{xx} \frac{ \gt (g_{zz}+\theta'^2) (\gx^2 + \B^2) + \gt \gx \( (\Ax)^2 + (\Ay)^2 \)}{\sqrt{d(z)}}.
\nonumber
\label{densities}
\end{eqnarray}
With straightforward algebra, we have checked that all the components of $\Theta^a_{~b}$ computed in this fashion agree exactly with those derived via the Noether procedure. Checking the $\Theta^i_{~z}$ components with $i=t,x,y$ requires the following derivatives:
\bea
\frac{\delta L}{\delta \partial_t \theta} & = & - \mathcal{N} \cos^3 \theta \, \gx^{1/2} \, \frac{ \gx \, \E \, \Ax}{\sqrt{d(z)}} \, \theta' \\
\frac{\delta L}{\delta \partial_x \theta} & = & - \mathcal{N} \cos^3 \theta \, \gx^{1/2} \, \frac{\gt \B \Ay - \gx \E \At}{\sqrt{d(z)}} \, \theta' \\
\frac{\delta L}{\delta \partial_y \theta} & = & + \mathcal{N} \cos^3 \theta \, \gx^{1/2} \, \frac{\gt \B \Ax}{\sqrt{d(z)}} \, \theta'.
\eea

\end{appendix}

\bibliographystyle{JHEP}
\bibliography{stressd7}

\end{document}